\documentclass[prx,twocolumn,showkeys,showpacs,superscriptaddress,nofootinbib,10pt
]{revtex4-1}
\usepackage{graphicx}  % needed for figures
\usepackage{grffile}
\usepackage{amsthm}
\usepackage{epsfig}
\usepackage{amsmath,amssymb,amsfonts,mathtools,bbm,MnSymbol,mathrsfs,xfrac}   % for math
\usepackage{longtable} %for tables
\usepackage{tabularx}
\usepackage[breaklinks=true,colorlinks=true,linkcolor=blue,urlcolor=blue,citecolor=blue]{hyperref}
\usepackage{comment}
\usepackage{color}
\usepackage[makeroom]{cancel}

\usepackage{soul}%for striking through text horizontally with \st{}
\renewcommand{\[}{\begin{equation}}
\renewcommand{\]}{\end{equation}}
\renewcommand{\Re}{\mathfrak{Re}}
\renewcommand{\Im}{\mathfrak{Im}}
\newcommand{\ket}[1]{|#1\rangle}
\newcommand{\bra}[1]{\langle#1|}
\newcommand{\braket}[2]{\langle#1|#2\rangle}
\newcommand{\pro}[2]{|#1\rangle\langle#2|}

\newcommand{\Eq}[1]{Eq.~\eqref{#1}}

\newcommand{\Pc}{\mathcal{P}}
\newcommand{\C}{{\mathcal{C}}}
\renewcommand{\P}{\hat{P}}

\newcommand{\bR}{{\mathbf{R}}}

\newcommand{\HS}{\mathcal{H}}

\newcommand{\bx}{{\boldsymbol{x}}}
\newcommand{\by}{{\boldsymbol{y}}}

\definecolor{mygray}{gray}{0.6}
\theoremstyle{definition}

\begin{document}

\title{%Everything eventually happens, but only halfway: a probability bound on extreme fluctuations in isolated quantum systems
Probabilistic bound on extreme fluctuations in isolated quantum systems}

\author{Joshua M. Deutsch}
\affiliation{Department of Physics, University of California, Santa Cruz, CA 95064, USA}
\author{Dominik \v{S}afr\'{a}nek}
\email{dsafrane@ucsc.edu}
\affiliation{SCIPP and Department of Physics, University of California, Santa Cruz, CA 95064, USA}
\author{Anthony Aguirre}
\affiliation{SCIPP and Department of Physics, University of California, Santa Cruz, CA 95064, USA}

\date{\today}

\begin{abstract}

We ask to what extent an isolated quantum system can eventually ``contract" to be contained within a given Hilbert subspace.  We do this by starting with an initial random state, considering the probability that all the particles will be measured in a fixed subspace, and maximizing this probability over all time. This is relevant, for example, in a cosmological context, which may have access to indefinite timescales. We find that when the subspace is much smaller than the entire space, this maximal probability goes to $1/2$ for real initial wave functions, and to $\pi^2/16$ when the initial wave function has been drawn from a complex ensemble. For example when starting in a real generic state, the chances of collapsing all particles into a small box will be less than but come arbitrarily close to $50\%$. This contraction corresponds to an entropy reduction by a factor of approximately two, thus bounding large downward fluctuations in entropy from generic initial states.
\end{abstract}

\maketitle

\section{Introduction}

It has long been known that although (suitably-defined) entropy tends to increase in a closed system, it can with exponential rarity also fluctuate downward.  Poincar\'e's recurrence theorem shows that under fairly general assumptions a classical system returns arbitrarily close to its initial state -- and hence entropy.  Fluctuation theorems have been proven quantifying the frequency of downward excursions within thermodynamics (e.g.~\cite{ sevick2008fluctuation}) and these excursions have even been observed in very small laboratory systems (e.g.~\cite{wang2002experimental}).

Although exponentially rare, these fluctuations can be of interest even in macroscopic systems in the context of cosmology and the theoretical study of eternal spacetimes.  Aguirre, Carroll and Johnson~\cite{aguirre2012out} studied several such contexts and processes including the formation of black (and white) holes in de Sitter or thermal anti de Sitter spacetime, thermal transitions in cosmological inflation, creation of a full Big-Bang universe from an eternal thermal bath, and formation of so-called ``Boltzmann Brains"\cite{albrecht2004can}. All of these are processes in which matter and energy spontaneously ``gather up" into a relatively small spatial region. At a classical level, as considered by Boltzmann, an extended and disordered system will spontaneously collapse into a very small volume, if given a long enough time to do so. The arguments of~\cite{aguirre2012out} show that this closely resembles the time-reverse of the dispersal of a localized collection of matter.  It is still a subtle problem, however, as to if and how this is possible in quantum mechanics. This is the subject of the present paper.

We consider {\em generic} many-body wave functions evolving under a time-reversal-symmetric Hamiltonian.
We consider two classes of initial wave functions $\ket{\psi(t=0)}$. The first is real at some time, say
$t=0$, corresponding to evolution with a $t \rightarrow -t$ symmetry.
In other words, all observables have an evolution that is even in time, which could be
of interest in some cosmological models possessing similar symmetries~\cite{aguirre2003inflation,carroll2004spontaneous,boyle2018c}. The other case is that of a wave function that is complex at $t=0$.

We study the probability that all particles are in some fixed relatively small region of
space. This probability will vary as the system evolves, and for the majority of the time it will be very small. However very
rarely, as elucidated below, this probability will become substantial. We argue that the maximum that this
probability attains is $1/2$ for the first class of wave functions $\ket{\psi(t=0)} \in \mathbb{R}$,
and $\pi^2/16$ for the second class of wave functions $\ket{\psi(t=0)} \in \mathbb{C}$.
Another way of putting this is that we can find a sufficiently large time where
the wave function will spontaneously localize into a given compact spatial region.
But only to a certain extent: in general for $\ket{\psi(t=0)} \in \mathbb{R}$ at least half of the probability will inevitably remain
spread throughout space (and this is $1-\pi^2/16$ for $\ket{\psi(t=0)} \in \mathbb{C}$). Of course, a wave function initially
confined to a local region will eventually (by a quantum recurrence
theorem) return to that local region. But this is not generic. In the generic case,
the probability of measuring a closed system
localized in some small subregion will {\em
eternally} be upper-bounded by one-half (for real initial wave functions). And the system becomes
arbitrarily close to attaining this bound during the course of its time evolution.

The above discussion is a brief description of our main result and to obtain this, we make a number of
technical assumptions that are specified in more detail in the following sections.
In Sec. \ref{sec:problemsetup} we define the problem and explain that for generic Hamiltonians, we can
transform the maximization over time to a maximization over phases. In Sec. \ref{sec:maximization},
we show that for finite dimensional Hilbert spaces, there is a way of simplifying this problem further by
performing a unitary transformation on the basis states of only the small region. This result allows
us, in Sec. \ref{sec:uncorr_mod}, to analyze various models of the wave function. These are models where the energy eigenstates, $\ket{E}$, have random
Gaussian statistics, and the wave function expansion coefficients, $\braket{E}{\psi}$ are similarly random.
If we further assume a finite but large Hilbert space, and a
subregion that is large but much smaller than the total size of accessible Hilbert space,
then we are able to obtain the maximum probabilities (e.g. 1/2) that were discussed above. We can also
predict the scaling of corrections to this maximum probability that are expected when the subregion
becomes larger.  Sec.~\ref{sec:numerics} analyzes the above models numerically and appears to confirm the
results predicted analytically. It also allows us to understand quantitatively how the maximum
probability varies as the subregion varies in dimension.

We can use the above results to understand these maximum probabilities in a more realistic system,
that of a quantum dilute gas above the ground state. It is shown in Appendix~\ref{sec:corr_sys}, that our main result, in the
appropriate regime specified there in detail, is still maintained.

We will argue in Sec.~\ref{sec-entropy} that with a suitable entropy definition, our main result indicates
that in a closed thermalized system, entropy can never decrease by
a factor of more than two (for $\ket{\psi(t=0)} \in \mathbb{R}$), unless the system at some point in its
past had an entropy lower than this.

\section{Problem setup}
\label{sec:problemsetup}

\begin{figure}[h]
\begin{center}
\includegraphics[width=1\hsize]{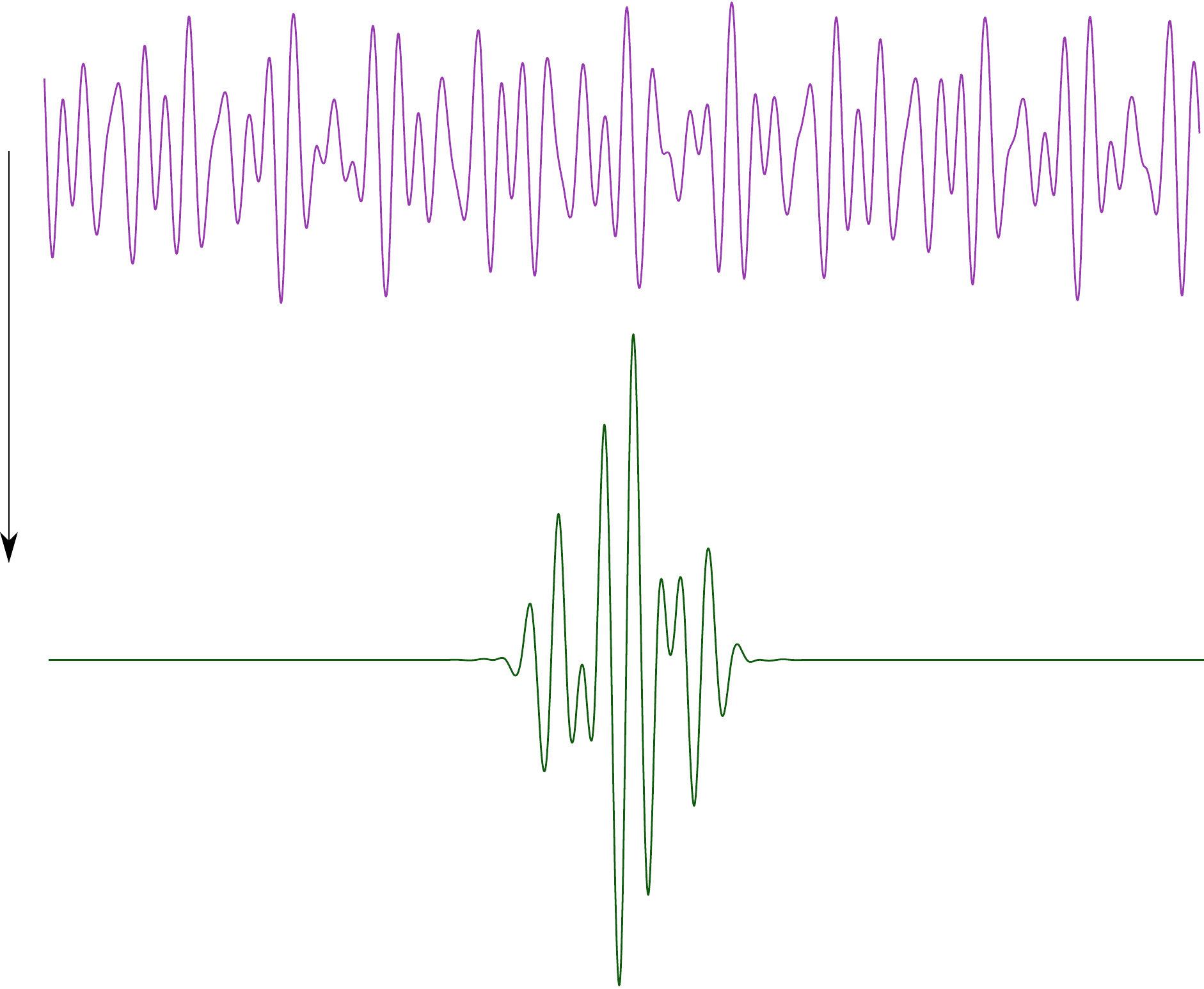}
\caption
{
A wave function starts off in a typical extended state and then evolves after some time to a function which is
localized to a much smaller region.
}
\label{fig:collapsing_psi}
\end{center}
\end{figure}

In order to apply this question to cosmology, it would make sense to include gravity, however a general quantum description of this is lacking
and we will therefore simplify the general problem by ignoring it and
assume that we have a large box in flat space-time of width $L$ housing our toy universe. The boundary conditions could be
periodic or there could be hard walls.  We further assume that the system is time-reversal invariant.
We assume that the wave function starts in some typical state, and therefore is one that
spans the entire box. We then ask if the wave function is ever, even after an arbitrarily long time,
able to evolve so that it is completely confined to a much smaller region, as shown
pictorially in Fig. \ref{fig:collapsing_psi}.

We start by discretizing space, which will allow us to more easily analyze this question quantitatively. Then, consider an arbitrary state vector $\ket{\psi_t}$ that can be written in an energy eigenbasis evolving in time
\[
\label{eq:def_psi_t}
\ket{\psi_t}  = \sum_E \braket{E}{\psi_t} \ket{E}= \sum_E c_E(t) \ket{E}
\]
where the coefficients $c_E(t) := \braket{E}{\psi}$ have a time dependence $c_E(t) = \exp(iEt) c_E(0) $.

Suppose the system has a fixed number
of particles $N_p$.  We are considering the system to be on a lattice and denote all of the coordinates
of the particles $\bx := (x_1,x_2,\cdots,x_{N_p})$ where each $x_i$ labels a lattice site. The particles could be indistinguishable, in which
case, the ordering does not matter. A positional basis state is denoted $\ket{\bx}$.

First, we ask whether particles that are scattered throughout the space can collapse into a single small region. Mathematically, starting with some random initial configuration, we ask whether at some time the wave function can be non-zero in an $M$-dimensional subspace $X$ of an $(N > M)$-dimensional configuration space.

If the energy eigenvalues $E$ are all incommensurate (irrationally related), as expected for a non-integrable system\cite{bohigas1984spectral,berry1987bakerian},
then the set of phase factors $\{\exp(i E t)\}_E$ will come arbitrarily close to any set of
complex unit magnitude numbers $\{z_E\}_E$ as time is varied. Therefore, rather than considering $\ket{\psi}$ as a function of time, we can write it as
\[
\label{eq:psi_of_zE}
\ket{\psi(\{z_E\}_E)} = \sum_E a_E z_E \ket{E},
\]
where $a_E := |c_E(0)|$, and the $z_E$ can each take any value on the complex unit circle. The sum is over all $N$ energy eigenvectors.

The wave function being contained within the region $X$ at some time in future is then equivalent to the existence of a set of $z_E$ values such that
\[
\braket{\bx}{\psi} = \sum_E a_E z_E \braket{\bx}{E} = 0\ \mathrm{for\ all}\ \bx\notin X.
\]
These are nonlinear complex equations in the $z_E$s, and we can take their real and
imaginary parts, giving $2(N-M)$ equations. The number of variables are the $N$ phase
angles. Therefore for generic values of the $\braket{\bx}{E}$
and the coefficients $a_E$, no solutions to this equation are possible unless $M \ge N/2$.

Therefore for $M \ll N$, we cannot expect a typical initial wave function to become completely
localized in this smaller region. But perhaps the system can come very close to being able to do this, leaving a tiny residue outside of $X$.
We can quantify this possibility
 by asking what is the {\em maximum probability}
of finding the system inside of region $X$, if it were to be measured.

\section{Maximization of probability}
\label{sec:maximization}

The probability of measuring the system to be in positional configuration $\bx$ is $p_\bx := |\braket{\psi_t}{\bx}|^2$.
We consider the probability of finding all of these particles within some region $X$ of Hilbert space with dimension $M$,
\[
\label{eq:p_X=sum_p_x}
p_X(t) := \sum_{\bx\in X} p_\bx = \sum_{\bx\in X} |\braket{\bx}{\psi_t}|^2,
\]
where we switch from the time-dependence to the $z_E$-dependence as
\[
\label{eq:p_x_z_E=sum_a_E_z_E}
p_\bx(\{z_E\}_E) = \big|\sum_E a_E z_E \braket{\bx}{E}\big|^2.
\]
We would like to find
\[
P_{\max} := \max_{0<t<\infty} p_X(t),
\]
which is given per the above arguments by
\[
\label{eq:P_max=max_z_E_sum_x_in_X_p_x}
P_{\max}  =  \max_{\{z_E\}_E} \sum_{\bx\in X} p_\bx(\{z_E\}_E).
\]

Finding this maximum exactly would in principle require solving $N$
equations for phases. Instead of doing so, we create an ansatz that
restricts our search space to a much smaller set of wave functions,
turning the problem into that of solving $M-1$ non-linear equations
with $M^2$ free parameters. The reason for this is two fold. First,
this will allow us to map this problem onto a related one, but with
$M=1$, which is helpful in understanding this problem analytically as done in Sec. \ref{sec:uncorr_mod}.
Second, it also gives rise to a useful method for numerically obtaining maximum probability
states. We were able to show in Appendix \ref{sec:UnitaryGlobalMaxEquiv} that the set of all
of the extrema of this ansatz are identical to the full set of extrema.
The point of the ansatz that we make is that it allows us to reduce the dimensionality
of the M-dimensional subspace down to one dimension. This becomes a much easier
problem to analyze.

The method proceeds as follows. First, we number basis vectors $\{\ket{\bx}\}_{\ket{\bx}\in X}$ as $\{\ket{\bx_i}\}_{i=1}^M$. Then we make a unitary transformation $U$ (represented by a unitary $M \times M$ matrix), changing the basis on subspace $X$ from $\{\ket{\bx_i}\}_{i=1}^M$ to $\{\ket{\by_i}\}_{i=1}^M$,
\[\label{eq:y_with_x}
\ket{\by_i} = \sum_{j=1}^M U_{ij} \ket{\bx_j}.
\]
For our argument, we will think of vectors $\ket{\by_i}$ as being fully dependent on $U$, $\ket{\by_i}=\ket{\by_i(U)}$, while $\ket{\bx_i}$ are fixed and given. Since both the $\ket{\bx_i}$ and the $\ket{\by_i}$ span the same subspace, they give rise to the same projector that projects onto the subspace,
\[
\label{eq:XProjectorEquivalence}
\hat{P}_X := \sum_{i=1}^M \pro{\bx_i}{\bx_i} = \sum_{i=1}^M \pro{\by_i}{\by_i}.
\]
We can then rewrite Eq.~\eqref{eq:p_X=sum_p_x} in terms this projector, and in terms of basis vectors $\ket{\by_i}$, as
\[\label{eq:p_X_in_P_X}
p_X =|\bra{\psi}\hat{P}_X\ket{\psi}|^2=\sum_{i=1}^M |\braket{\by_i}{\psi}|^2.
\]
(Let us recall that $\ket{\psi}=
\ket{\psi(\{z_E\}_E)}$
as per Eq.~\eqref{eq:psi_of_zE}.)
We now use our freedom to choose $U$ to find a particularly simple basis in which
\[
\label{eq:psi_y=0}
\braket{\by_i(U)}{\psi} = 0~\mathrm{for}~i=2,\cdots,M .
\]
An arbitrary $M\times M$ unitary matrix has $M^2$ independent real parameters, so it seems reasonable that one should exist imposing these $M-1$ conditions.  And indeed it can be explicitly constructed:
 it is the transformation that aligns the new basis in such a way that $\ket{\by_1}$ points in the same direction as the projection of the wave function $\ket{\psi}$ onto the subspace, i.e., this unitary transformation sets
 \[
 \label{eq:y1=proj_psi}
 \ket{\by_1}:=\frac{\hat{P}_X\ket{\psi}}{||\hat{P}_X\ket{\psi}||},
\]
and the set of conditions, Eq.~\eqref{eq:psi_y=0}, is then trivially satisfied. Conversely, if the set of conditions is satisfied, then it must be that $\ket{\by_1}:=\frac{\hat{P}_X\ket{\psi}}{||\hat{P}_X\ket{\psi}||}$.

With this choice of unitary transformation, and use of Eq.~\eqref{eq:psi_of_zE} we can write Eq.~\eqref{eq:p_X_in_P_X} as
\[
\label{eq:pX=sum_y1_E_sq}
p_X =
 |\braket{\by_1(U)}{\psi}|^2 = \big|\sum_E a_E z_E\braket{\by_1(U)}{E}\big|^2 .
\]

Clearly, this expression achieves its maximum when the phases are given the values $\tilde z_E$ that maximize every term in the energy sum,
\[
\label{eq:maxzE}
\tilde{z}_E := \frac{\braket{E}{\by_1(U)}}{|\braket{E}{\by_1(U)}|}.
\]
It should be noted that no solution to these equations need exist because they are implicitly nonlinear. The reason is that desired phases $\tilde{z}_E$ depend on the choice of basis vector $\ket{\by_1(U)}$, which, of course,  depends on the $U$. But $\ket{\by_1}$ (or equivalently $U$) is constructed through Eq. \eqref{eq:y1=proj_psi}, and that depends on $\ket{\psi}$. But $\ket{\psi}$ depends on the desired phases $\tilde{z}_E$. These are therefore a nonlinear set of coupled equations. This concern is compounded by the fact that Eq. \eqref{eq:maxzE} requires every term in Eq. \eqref{eq:pX=sum_y1_E_sq} is individually maximized and it is not clear that this is possible to achieve. However  we now show that these equations are indeed general solutions for extrema of $p_X$, Eq. \eqref{eq:p_x_z_E=sum_a_E_z_E}. 

Using Eqs. \eqref{eq:maxzE}, \eqref{eq:y1=proj_psi}, and \eqref{eq:psi_of_zE}, we show in
Appendix \ref{sec:UnitaryGlobalMaxEquiv} that
\[
\label{eq:z_EproptosumJz}
   \arg(\tilde{z}_E) = \arg(\sum_{E'} J_{E E'} \tilde{z}_{E'})
\]
where $\arg$ denotes the complex phase, $z=|z|e^{i \arg{z}}$, and
\[
\label{eq:JEE=sumaaExxE}
J_{EE'} := \sum_{x\in X} a_E a_{E'}\braket{E}{x}\braket{x}{E'}.
\]
can be interpreted as the coupling between spins, as shown in Appendix \ref{sec:rare_fluctuations}.
It is further shown in Appendix \ref{sec:UnitaryGlobalMaxEquiv} that Eq. \eqref{eq:z_EproptosumJz} is the same as obtained by directly
extremizing the probability given in Eq. \eqref{eq:P_max=max_z_E_sum_x_in_X_p_x}. Denoting $\ket{\tilde{\by}_1}$ as a transformed
basis state corresponding to the global maximum, we have
\[
\label{eq:pX=sum_aE_Ey}
\tilde{P}_{\max} = \bigg[\sum_E a_E |\braket{\tilde{\by}_1}{E}|\bigg]^2.
\]
This shows that
this unitary ansatz gives the general solution for extrema. This means that we can always find a basis
where we need consider only the wave function at a single point in Hilbert space
rather than on an $M$-dimensional subspace. This property is very useful in
understanding analytically $P_{max}$ as we shall now explain.

\section{Uncorrelated Model}
\label{sec:uncorr_mod}

\subsection{Constant variance}
\label{sec:uncorr_mod_constant}

Given the above solution for the probability maximum, we turn to evaluation of this maximum under different models for the amplitudes $a_E$.
We consider that our states live in a finite $N$-dimensional Hilbert space, and that the Hamiltonian implies time symmetry, allowing the choice of a real eigenbasis. There are $N$ energy eigenvectors
$\braket{E}{\bx_i}$, $\bx_i\in\{\ket{\bx_i}\}_{i=1}^M$, and we can think of this as a matrix of $N$ rows, and $M$ columns. Here there are $N$ possible values of $E$ and $M$ values of $\bx$. Any two distinct columns of this matrix are orthonormal.
As our first model, we choose the set of coefficients $\{a_E\}_E$ to
be uniform subject to the constraint of unitarity $\sum_E |a_E|^2 = 1$. This can be seen to be equivalent for almost all
purposes, to independent Gaussian variables. This is identical to the reasoning behind the
equivalence between microcanonical and canonical ensembles. For example
for an ideal gas with energy conservation, when the number of particles is very large, we have $\sum
p_i^2$ is constant ($p_i$ are momenta). The $p_i$'s are chosen uniformly subject to this constraint and
this microcanonical distribution becomes equivalent to the Maxwell-Boltzmann distribution.
This appears to have first been understood formally by Lax~\cite{lax1955relation}. Higher order energy-energy
correlations become negligibly small and individual momenta obey Gaussian statistics.

Therefore the absolute value of independent Gaussian real random numbers are distributed as
\[
\label{eq:PaE-exp-asq}
P(a_E) d a_E \propto \exp(-a_E^2/2\sigma_a^2) d a_E .
\]
From unitarity, $\sum_E a_E^2 = 1$. We start by considering the variances of the $a_E$ to all be
the same, which gives $\langle\sum_E a_E^2\rangle=N\langle a_E^2\rangle=1$, that is $\sigma_a^2 = 1/N$.

We are taking a model where matrix elements $\langle \bx | E\rangle$ are independent random Gaussian variables. Although this is not generally true for real physical systems, this simplified model will be then useful in analyzing the more realistic situation of an interacting gas, described in Appendix~\ref{sec:corr_sys}.

To keep the notation compact, we also identify $\by_1:= \tilde{\by}_1$, and $P_{max}:= \tilde P_{max}$ from Eq.~\eqref{eq:pX=sum_aE_Ey}. Consider
\[\label{eq:pX=sum_aE_Ey2}
\sqrt{P_{max}} = \sum_E a_E |\braket{E}{\by_1}|,
\]
and write
\[
\label{eq:y1=sum_gamma}
\ket{\by_1} = \sum_{j=1}^M U_{1j} \ket{\bx_j} ,
\]
where $U_{1j}$ are the appropriate matrix elements of the unitary transformation connecting the $\bx$
and $\by$ bases.

Because the elements of unitary transformation $U_{1j}$ are complex numbers which are expected to have uniformly distributed complex phases, since $\phi_E := \braket{E}{\by_1}$ is a sum of those random numbers by Eq.~\eqref{eq:y1=sum_gamma}, it must be distributed as a Gaussian complex variable, i.e.,
\[
\label{eq:PphiDistro}
\begin{split}
&P(\phi_E) d \mathrm{Re}(\phi_E) d \mathrm{Im}(\phi_E) \propto\\
&\exp(-|\phi_E|^2/2\sigma_E^2) d \mathrm{Re}(\phi_E) d \mathrm{Im}(\phi_E).
\end{split}
\]
$\sigma_E^2$ is again obtained through unitarity $\langle \sum_E |\phi_E|^2\rangle = N\langle |\phi_E|^2\rangle = N \langle\Re(\phi_E)^2 + \Im(\phi_E)^2\rangle =1$, so that $\langle \Re(\phi_E)^2\rangle =\langle \Im(\phi_E)^2\rangle = \sigma_E^2 = 1/2N$.
Eq.~\eqref{eq:pX=sum_aE_Ey2} involves a sum over a large number of independent variables and so is
self-averaging. Therefore, we can take Eq.~\eqref{eq:pX=sum_aE_Ey2} and take its average, which must give the same answer as without averaging,
\[
\label{eq:sqrtp_X=1sqrt2}
\sqrt{P_{max}} = \sum_E a_E |\phi_E| = \langle\sum_E a_E |\phi_E|\rangle.
\]
As we will show shortly, in the limit of $M^2 \ll N$, the correlations between $a_E$'s and $\phi_E$'s are so weak, that the result we obtain for that case is almost the same as in the case when $a_E$'s and $\phi_E$'s are uncorrelated. We will therefore consider  $a_E$'s and $\phi_E$'s to be independent random variables, which gives,
\[
\begin{split}
\label{eq:sqrtp_X=1sqrt2b}
\sqrt{P_{max}} &= \langle\sum_E a_E |\phi_E|\rangle = N \langle a_E\rangle \langle |\phi_E|\rangle \\
&=N\ \! \frac{\sqrt{2}}{\sqrt{\pi N}}\ \! \frac{\sqrt{\pi}}{2\sqrt{N}}=\frac{1}{\sqrt{2}}.
\end{split}
\]

Since $\sqrt{P_{max}}$ is self-averaging, also $P_{max}$ is also self-averaging, which gives the final result,
\[
\label{eq:p_X=1_ov_2}
P_{max} = \langle[\sum_E a_E |\phi_E|]^2\rangle = \frac{1}{2},~\mathrm{for}~M^2\ll N.
\]

If instead of drawing the coefficients $a_E$ randomly from a Gaussian distribution of real numbers, we choose them from a complex Gaussian ensemble, similar to~\Eq{eq:PphiDistro},
this changes the limiting value to $P_{max} = \pi^2/16\approx 0.617$.

Now, let us take a look at the validity of the assumption made above Eq.~\eqref{eq:sqrtp_X=1sqrt2b}, i.e., that $a_E$'s and $\phi_E$'s are so weakly correlated that they give the same result for the maximum as independent random variables would give.

If we transform the $\bx$ basis into the final $\by$ basis according to Eq.~\eqref{eq:y1=sum_gamma}, the random numbers will become correlated because of the maximization procedure. However in the
limit where $M^2 \ll N$, this maximization can only influence $M^2$ degrees of freedom and therefore
has a negligible effect on the independence of the different terms $\braket{E}{y_1}$ used in \Eq{eq:pX=sum_aE_Ey}.

We show that explicitly by showing that even when we vary $U_{1j}$, the final maximum does not change much, and is more or less equal to $\frac{1}{\sqrt{2}}$ as given by Eq.~\eqref{eq:sqrtp_X=1sqrt2b}. In other words, we will study variation in the function
\[
\sqrt{p_X}(\{U_{1j}\}):= \sum_E a_E |\sum_{j=1}^MU_{1j}\braket{E}{\bx_j}|,
\]
constructed by substituting Eq.~\eqref{eq:y1=sum_gamma}
into Eq.~\eqref{eq:pX=sum_aE_Ey2}. $U_{1j}$ in the above function introduces correlations between $a_E$'s and $\phi_E$'s, so if we are able to show that this function does not change much when we vary $U_{1j}$, then we can conclude that correlations between $a_E$'s and $\phi_E$'s do not really matter, and they can be considered uncorrelated.

To estimate how much $\sqrt{p_X}$ changes, we can differentiate $\sqrt{p_X}$ with respect to the $U_{1j}$'s (with
$U_{1j}^*$ being an independent variable),
\[
\label{eq:part_sqrt_p_x=sum_E_a_E}
\frac{\partial\sqrt{p_X}}{\partial U_{1j}} = \sum_E a_E \braket{E}{\bx_j} \frac{1}{2} e^{- i \arg(\sum_j U_{1j} \braket{E}{\bx_j})}
\]
Each term in the summation is of order $N^{-1/2}N^{-1/2}$ and the phase angles in the exponential will fluctuate randomly as a function of $E$, making the sign of
each term in the sum random. The addition of $N$ such terms leads to an answer of magnitude
$\frac{\partial\sqrt{p_X}}{\partial U_{1j}}\sim N^{-1/2}$ with a variable sign that depends on the values of the $U_{1j}$'s.

Note that the above argument will still hold if there are local correlations between neighboring
$\bx$'s for both $\braket{E}{\bx_j}$.

Each matrix element $U_{ij}$ of a unitary transformation has $|U_{ij}| \le 1$. Therefore the maximum
deviation of  $\sqrt{p_X}$ from its mean can be estimated by
\[
\Delta \sqrt{p_X} \le \max_{\{U_{1j}\}} \bigg\rvert\sum_{j=1}^M \frac{\partial\sqrt{p_X}}{\partial U_{1j}} \Delta U_{1j}\bigg\rvert,
\]
where $\Delta$ represents the difference between an arbitrary initial value of $U_{1j}^{(0)}$ and its
final value, $\Delta U_{1j} = U_{1j} - U_{1j}^{(0)}$, and the maximum goes over all combinations $\{U_{1j}\} := \{U_{11},\dots,U_{1M}\}$.
Although the sign of the partial derivatives varies, with $M$ separate $U_{1j}$'s we expect that we can
choose values of the $U_{1j}$'s to
make every term in the sum positive. (If not, $\Delta p_X$ would be even less than this estimate.)
Therefore
\[
\label{eq:DeltaSqrtP_X=O_N_over_Msq}
\Delta \sqrt{P_{max}} \sim  O\bigg(\frac{M}{\sqrt{N}}\bigg) = O\bigg(\bigg(\frac{N}{M^2}\bigg)^{-\frac{1}{2}}\bigg).
\]
In the limit of large $N/M^2$, the difference between the typical values of $p_X$ and its maximum
vanishes. Note also that this implies that $P_{max}$ depends only on the combination $N/M^2$.
\begin{comment}
We can see that we would expect this to be true more generally
because the $M\times M$ unitary matrix has $M^2$ independent
degrees of freedom. Therefore one would expect that we can write
\[
\label{eq:P_XN_o_Msq}
p_X(N,M) = P\bigg(\frac{N}{M^2}\bigg).
\]
\end{comment}
Therefore one would expect that we can write
\[
\label{eq:P_XN_o_Msq}
p_X(N,M) = P\bigg(\frac{N}{M^2}\bigg).
\]
We will be see numerical confirmation of this scaling prediction in Sec.~\ref{sec:numerics}.

\subsection{General variance}
\label{subsec:GeneralVariance}

Now we extend this analysis to the situation where the coefficients $a_E$ are not statistically
identical but have a variance that depends smoothly on $E$. That is $\langle
a_E^2\rangle = \sigma_a^2(E)$, where the latter is some smoothly varying function. As explained at
the beginning of this section, a uniform choice of $a_E$'s with the constraint of unitarity is equivalent for almost all purposes to
that independent Gaussian variables. We are now weighting the $a_E$'s by additional independent
Gaussian probability factors. This implies the $a_E$'s for large $N$ are statistically independent.

We still need to maximize $p_X$ in accordance with \Eq{eq:pX=sum_aE_Ey}. Assuming again no correlation between the $a_E$ and the $\phi_E$, \Eq{eq:sqrtp_X=1sqrt2} becomes
\[
\sqrt{P_{max}} = \langle\sum_E a_E |\phi_E|\rangle = \sum_E \langle a_E\rangle \langle |\phi_E|\rangle.
\]
Still assuming Gaussian statistics for the coefficients in these sums, and following similar logic
to the uncorrelated case, we have
\[
\label{eq:pX=1_2sum_sigma_a_sigma_phi}
\sqrt{p_X} = \frac{1}{\sqrt{2}} \sum_E \sigma_a(E) \sigma_\phi(E),
\]
where we have defined $\sigma_\phi(E):= \langle |\phi_E|^2\rangle^{1/2}$.

In choosing the basis vector $\ket{\by_1}$ \Eq{eq:y1=sum_gamma} has $M$ parameters $\{U_{1i}\}$
that can be varied.
Thus we can use these degrees of freedom to choose the variances $\sigma_\phi(E)$ by changing the basis.
For sufficiently large $M$, we should be able
to maximize $p_X$ with respect to $\sigma_\phi(E)$, but with the constraint of
unitarity, which means that
\[
1 = \langle\braket{\by_1}{\by_1}\rangle = \langle\sum_E \braket{\by_1}{E}\braket{E}{\by_1}\rangle = \langle\sum_E |\phi_E|^2\rangle=
\sum_E\sigma_E^2.
\]
Adding this in with a Lagrange multiplier $\lambda$, we are maximizing
\[
L = \sum_E \sigma_a(E) \sigma_\phi(E) + \lambda\sum_E\sigma_\phi(E)^2
\]
with respect to the $\sigma_\phi(E)$. This gives $\sigma_\phi(E) = a_E$. Substituting this into
\Eq{eq:pX=1_2sum_sigma_a_sigma_phi} gives $p_X = 1/2$ as was found in the previous section.
And similarly, if $\braket{\psi}{E}$ is drawn from a Gaussian complex ensemble, $p_X = \pi^2/16$.

The above analysis will only work if $M$ is sufficiently
large and $a_E$ does not vary strongly with $E$. In the opposite limit where there is a strong variation
of $a_E$ with $E$, and $M$ is small, we cannot perform a maximization without adding additional
constraints and the answer is expected to be smaller.

As an example with quickly-varying $a_E$, consider a model with an energy cutoff $E_c$, such that $\sigma_a(E)$ is constant,  below $E_C$
and  $a_E=0$ above it. Correspondingly we denote $N_C$ as the number of non-zero $a_E$ terms.
We can repeat the same steps leading to \Eq{eq:DeltaSqrtP_X=O_N_over_Msq}. Now $\sigma_a(E) = 1/\sqrt{N_C}$  for
$E < E_C$, and there are $N_C$ non-zero terms in \Eq{eq:part_sqrt_p_x=sum_E_a_E}, leading to the
same order of fluctuation for this partial derivative. Therefore we still expect that the maximum
fluctuation of $p_X$ from its mean will still be $O(M/\sqrt{N})$, and, in this limit this is taken to
be small. Therefore we can estimate $P_{max}$ by taking its typical value as was done before. Repeating
the same analysis as leading to  \Eq{eq:p_X=1_ov_2}, now we obtain
\[
\label{eq:p_X=Nc_ov_2N}
P_{max} = \frac{N_c}{2 N},~\mathrm{for}~M^2\ll N.
\]

We also performed an analysis with correlated systems and initial real wave-functions, which can be found in Appendix~\ref{sec:corr_sys}. There we studied two regimes for subsystem $X$: Regions much smaller than a cubical region of width given by the thermal wavelength, and regions much larger than the thermal wavelength but still significantly smaller than the full system. For small regions we found
\[
\sqrt{P_{max}}= \frac{2}{\pi} \bigg(\frac{2 l}{\lambda_T}\bigg)^{\frac{d N_p}{2}}.
\]
Therefore in this limit, $P_{max}$ is proportional to the volume of $X$, independent
of system size, but dependent on temperature $T$, and the number of particles $N_p$.
For larger regions we found a result identical to that for uncorrelated system,
\[
P_{max}= \frac{1}{2}.
\]

\section{Numerics}
\label{sec:numerics}

\begin{figure}[htp]
\begin{center}
   \includegraphics[width=1\hsize]{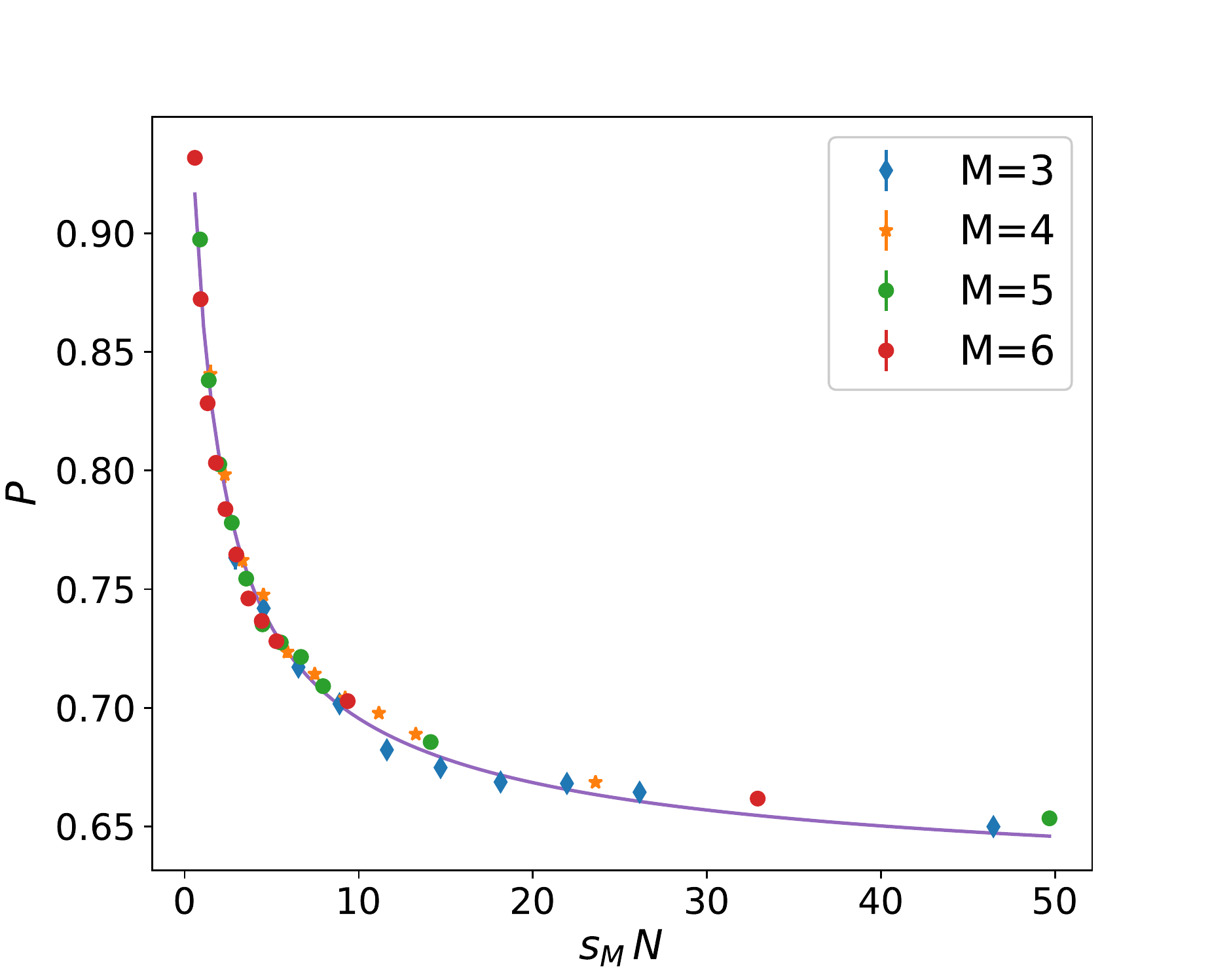}\\
   ~~~~~(a)\\
   \includegraphics[width=1\hsize]{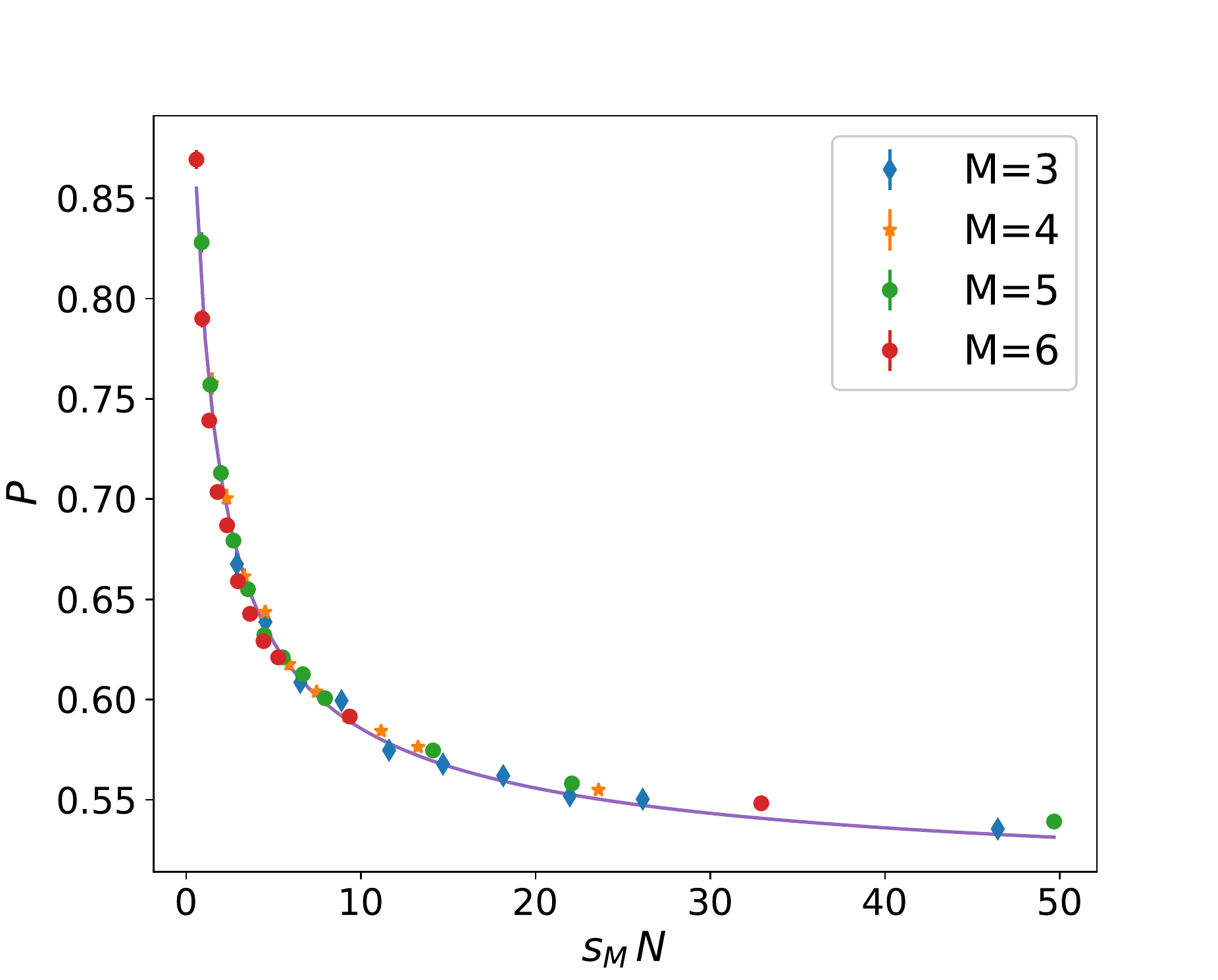}\\
   ~~~~~(b)\\
\caption
{
The maximal probability $p_X$ computed for Hilbert spaces of varying dimensions $N$ and subspaces $X$ of dimensions $M$, in the uncorrelated constant variance model,
$\sigma_a(E)=\mathrm{const.}$ as described in Sec.~\ref{sec:uncorr_mod_constant}. The horizontal axis rescaled by a factor plot $s_M$ in Eqs.~\eqref{eq:PNM=PsMN} and \eqref{eq:SsM=1overMsq}.
The number of separate random
instances for each data point is $300$ and the error bars for each point are also shown.
   (a) Results for the complex ensemble. (b) The results for the real ensemble.
   In both cases, the solid line is a fit to the function given in Eq.~\eqref{eq:PNxFitFunction}.
}
\label{fig:P_vs_N}
\end{center}
\end{figure}

\begin{figure}[htp]
\begin{center}
\includegraphics[width=1\hsize]{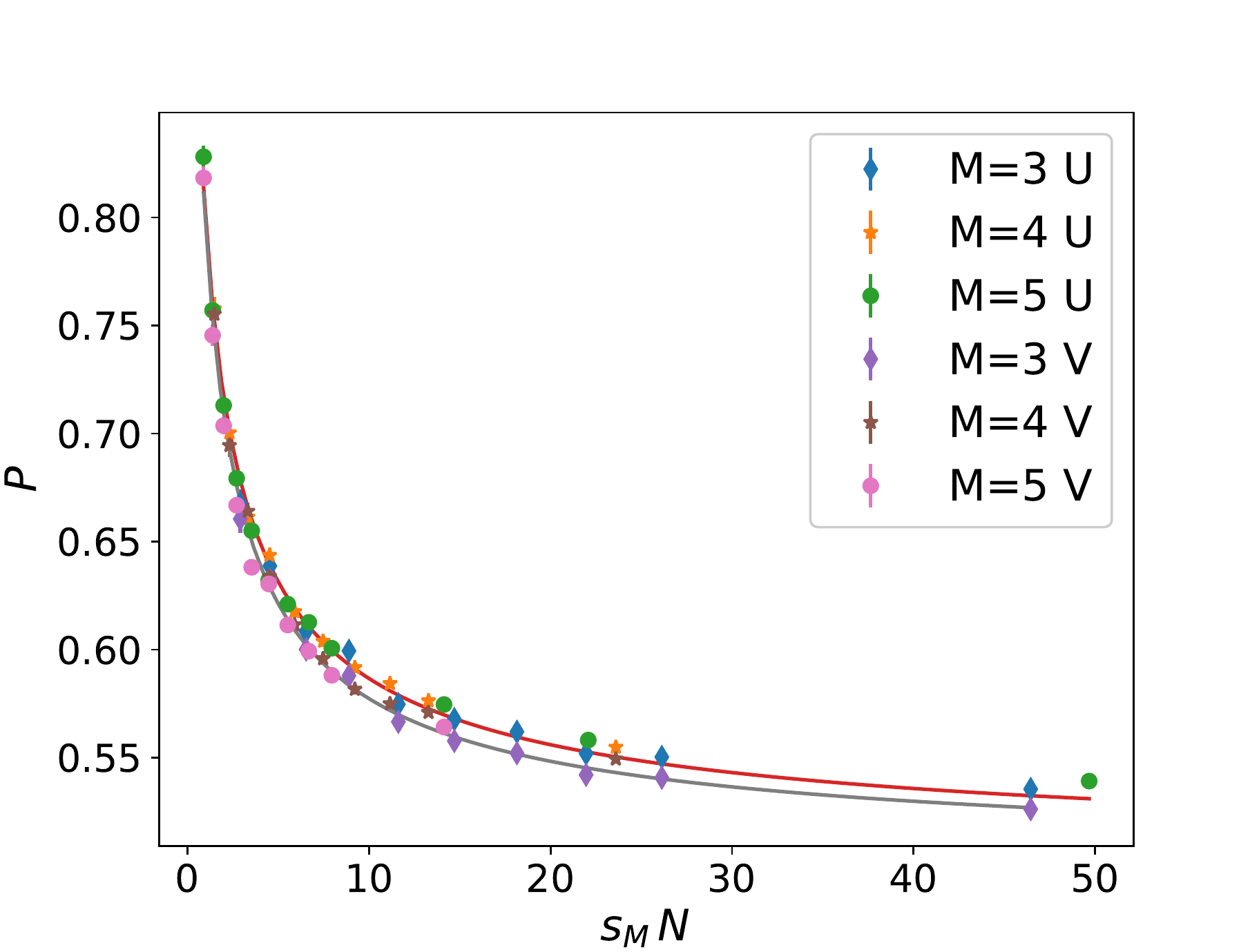}
\caption
{
The maximal probability $p_X$ computed for a range of different dimensions of Hilbert space $N$ and for the
subspace $X$ dimensions $M=3$, $4$, and $5$, for two different models for the coefficients in the spectral
expansion of the wave function $a_E$. The ``U'' denotes data from the uncorrelated uniform (that is constant)
variance model $\sigma_a(E)=\mathrm{const.}$,
and the  ``V'' represent a model with variance that varies exponentially with energy, Eq.~\eqref{eq:correlated_with_E}.
The fits and scaling are performed in the same way as in in Fig. \ref{fig:P_vs_N}
}
\label{fig:comp_const_exp_M3}
\end{center}
\end{figure}

\begin{figure}[htp]
\begin{center}
\includegraphics[width=1\hsize]{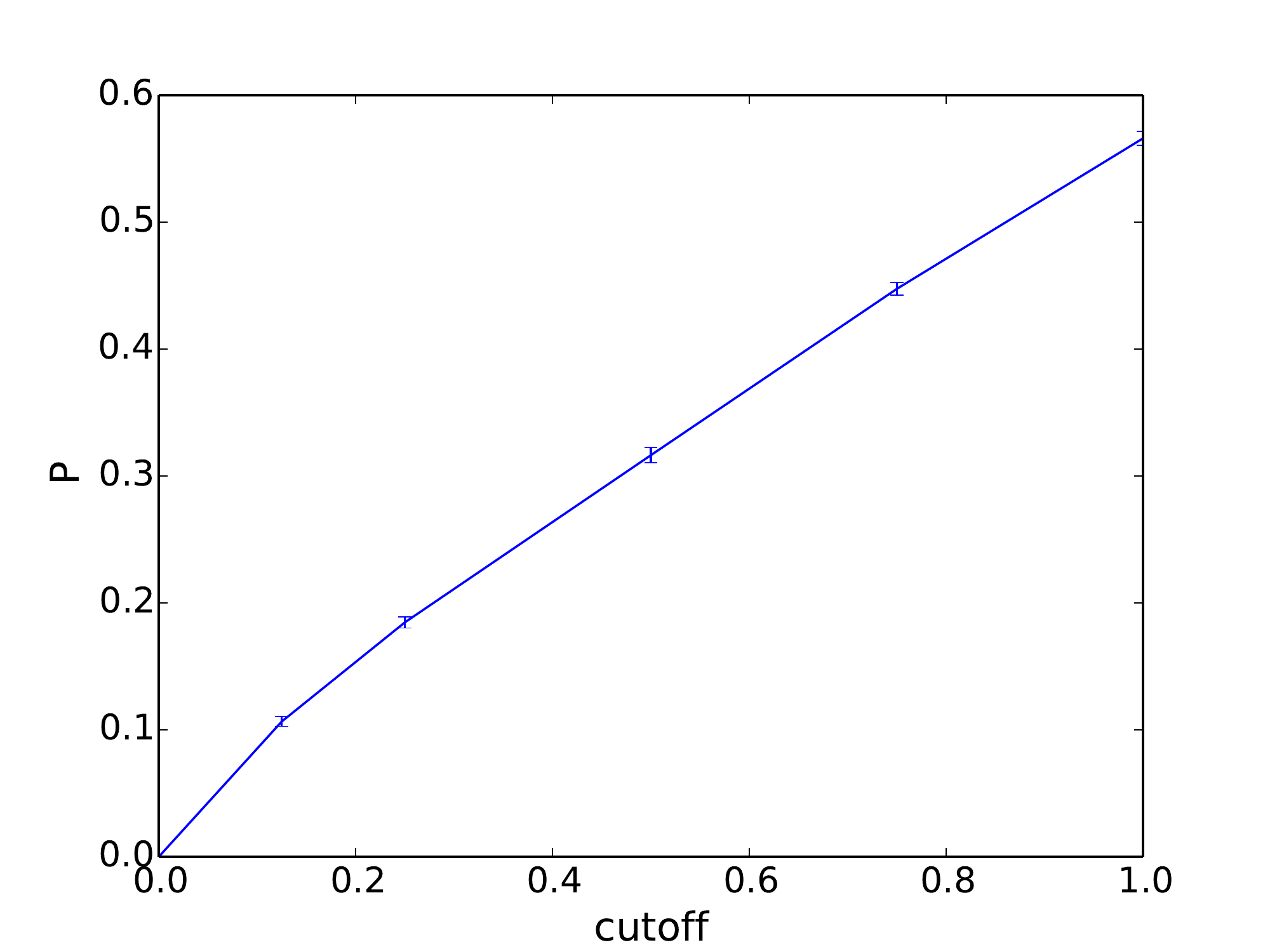}
\caption
{
The maximal probability $p_X$ as plotted as a function of the fraction of non-zero energy coefficients $a_E$, where the non-zero coefficients are taken from the uncorrelated constant variance model $\sigma_a(E)=\mathrm{const.}$ The cutoff parameter
on the horizontal axis is the ratio of the total number of nonzero coefficients to the dimension of the Hilbert space $N$.
The data are for $N=81$ and $M=3$.
}
\label{fig:P_cutoff}
\end{center}
\end{figure}

We now perform numerical computations to compare with our analytical predictions.
%that the solution that we found works and is a maximum.

Starting with the uncorrelated constant variance model of Sec.~\ref{sec:uncorr_mod_constant}, we maximize the probability $p_X$ over
the space of all $M\times M$ unitary transformations, transforming the $x$ to the $y$ basis
and choosing the phases $z_E$ in accordance with Eq.~\eqref{eq:maxzE}. For
$M=2,3,4,5,6$, after minimization we verified that it satisfied Eq.~\eqref{eq:psi_y=0}. To do the unitary maximization, instead of constructing an $M\times M$ unitary matrix, we did the maximization in steps. In one step, we maximized by choosing two $\bx$ values randomly, and constructing $2\times 2$ unitary transformations in that subspace. $p_X$ was maximized over those $2\times 2$ transformations, and then the process was repeated.

To test that the solution that we found is indeed a local maximum, we study the second derivative of $p_X$. This can be calculated as
\[
\frac{\partial^2 p_X}{\partial\theta_E\theta_{E'}} = \mathrm{Re}\Big( -2 \Big(\sum_{E"}
t^{(1)}_{E"}\Big)t^{*(1)}_E\delta_{EE'}+2\sum_{j=1}^M t^{(j)}_E t^{*(j)}_{E'}\Big),
\]
where
\[
t^{(j)}_E := a_E \tilde{z}_E \braket{\by_j}{E}.
\]
We numerically found that the largest non-zero eigenvalue is always negative in the parameter range that we discuss below. The solution is therefore stable.

To provide an additional check, we developed a different method (see Appendix~\ref{IterativeProcedureForMax} in which we started the procedure at a number of initial random phases (e.g. $1200$)), and used Eq.~\eqref{eq:y1=proj_psi} to determine $\ket{\by_1}$ for a fixed choice of phases. Then new phase angles $\{\theta_E\}_E$ were computed through Eq.~\eqref{eq:maxzE} and this process was repeated until the $L2$ norm of first derivatives $\frac{\partial{p_X}}{\partial\theta_E}$ had a magnitude less than $10^{-20}$. By starting this procedure from different initial random phases, we determined all of the local maxima and determined the global maximum. 

We tested that the solution using this unitary method finds a global maximum by running the maximization ($N_s = 25$) times and comparing it with the second method just described. For $m=4$ and $N=16$, and $m=5$ and $N=25$, $P_{\max}$
found by the maximization procedure over unitary transformations always differed relatively by less than $7\times 10^{-4}$
from $P_{\max}$ obtained by phase maximization, and spot checks showed that the same maxima were being found. Therefore the two methods were found to be in agreement.

Having verified that the solution is a maximum, we then computed
how $p_X$ varies with $N$ and $M$ and tested to see if it obeyed
the scaling prediction of \Eq{eq:P_XN_o_Msq}. We wrote this maximal probability in scaling form
\[
\label{eq:PNM=PsMN}
   P(N,M) = P(s_M N)
\]
where $s_M$ is a scale factor that for large $M$, should be $1/M^2$ and thus we expect
finite size corrections, the first term of which we take to be $1/M^3$. Therefore we
tried to fit using
\[
   \label{eq:SsM=1overMsq}
   s_M = \frac{1}{M^2} +\frac{b}{M^3}
\]
where $b$ is a constant that is chosen to obtain the best collapse. Numerically we find we get the
best collapse where $b \approx 1.9$, although the presence of this term is not a large effect.
The data scales quite nicely
as shown with the numerical results for both the complex, (a), and real (b) ensembles.
Fig.~\ref{fig:P_vs_N} shows the results for
computing $p_X$ for a range of values. Each point represents the average of the  maxima found for $300$ $N\times M$ energy eigenvector matrices, that is, the orthonormal matrix elements $\braket{E}{\bx}$. The $y$-axis represents the average of $p_X$ and this average's associated error bar.  The $x$-axis represents $s_M N$. As can be seen, good collapse of the data is achieved with this choice of scaling. For large $N/M^2$, $p_X$ appears to be converging slowly towards $1/2$.  \Eq{eq:DeltaSqrtP_X=O_N_over_Msq} explains the slow
convergence to $P_{max}=1/2$ for large $N/M^2$ in Fig.~\ref{fig:P_vs_N}.

The scaled data were also fitted according to the following functional form which however does not have any theoretical justification
\[
\label{eq:PNxFitFunction}
P_N(x) = 1-(1-P_\infty)\exp(-b/x^c)
\]
Here $P_\infty$ are the asymptotic values of the maximum probabilities predicted here theoretically,
\[
   P_\infty = \lim_{x\rightarrow\infty} P(x) =
   \begin{cases}
      \frac{\pi^2}{16} &\text{~for } \{a_E\}_E \in \mathbb{C},\\
      \frac{1}{2} &\text{~for } \{a_E\}_E \in \mathbb{R},
    \end{cases}
\]

\[
   \begin{split}
       b = 0.87, c = 0.66   &\ \ \text{~for } \{a_E\}_E \in \mathbb{R},\\
       b = 1.07, c = 0.67    &\ \  \text{~for } \{a_E\}_E \in \mathbb{C}.
   \end{split}
\]

We also test what happens if the coefficients $a_E$ are not statistically identical as was
analyzed in Sec.~\ref{subsec:GeneralVariance}. (All results that follow are shown for the real
ensemble).  We first choose
\[\label{eq:correlated_with_E}
\sigma_a(E) \propto \exp[-n_E/(2N)],
\]
where $n_E$ is an integer index corresponding to energy eigenvalue $E$. Because the time dependence
is no longer present, the values of $E$ are irrelevant, and the ordering of the different energies
is arbitrary.  Fig.~\ref{fig:comp_const_exp_M3} plots $p_X$ as a function of $N$ for both the
uniform (constant) case (U), and the above variance (V). The results are quite close to each
other and appear to be slowly converging to $1/2$ for large $N$. This what we expect when the
$\sigma_a(E)$ do not vary much with $E$.

However for faster decay rates in $\sigma_a(E)$, as expected in realistic system with finite energy, the results vary significantly. To analyze this more quantitatively, we choose a model with an energy cutoff as was done at the end of Sec.~\ref{subsec:GeneralVariance}. We define a cutoff parameter as the ratio of the number of nonzero $a_E$ coefficients
to $N$. We run the model for  $M=3$ and $N=81$, employing different cutoff parameters ranging from $0$ to $1$.
The results are shown in Fig.~\ref{fig:P_cutoff}. Here we see an almost linear dependence on the cutoff, and the value when the cutoff is 1
is greater than $1/2$, as expected from Fig.~\ref{fig:P_vs_N}.

Larger Hilbert spaces and more realistic systems, involving lattice models of particles have also
been analyzed numerically~\cite{faiez2019extreme} to find the probability maxima and also examine
the entropy.

\section{Entropy}
\label{sec-entropy}

When we compare the behavior of a quantum to an analogous classical system, there are some well known important differences,
such as the quantization of energy levels, appearance of superfluidity, macroscopic and microscopic interference effects.
Adding to this is the behavior of
very rare fluctuations in isolated systems: in this section we show that in terms of entropy decrease these fluctuations occur rather differently in quantum systems versus in classical ones. (See Appendix~\ref{sec:rare_fluctuations} for more detail regarding the form of the fluctuations.)

An isolated classical gas will always undergo rare but significant
reductions in entropy. If the gas is close to being ideal, these
can be quantified as follows. If we denote the number of (monatomic)
particles as $N$, and their mass $m$, the thermodynamic entropy  as a function
of the total energy $E$, and volume $V$ is~\cite{ma1985statistical}
\[
\label{eq:ClassicalGasEntropy}
S_I(E,V) = N \bigg(\frac{3}{2}\ln\bigg(\frac{4\pi m E}{3Nh^3}\bigg) + \ln\bigg(\frac{V}{N}\bigg) + \frac{5}{2} + O\bigg(\frac{\ln N}{N}\bigg)\bigg).
\]
This can be rewritten as
\[
\label{eq:ClassicalGasEntropy2}
S_I(E,V) = N \bigg(\ln\bigg(\frac{(V/N)}{\lambda_T^3}\bigg) + O\bigg(\frac{\ln N}{N}\bigg)\bigg).
\]
where $\lambda_T$ is, to within trivial constants, the thermal wavelength (see \Eq{eq:lambda_TDef}).

Inasmuch as entropy can be defined out of thermal equilibrium, when the system spontaneously contracts to a ball of much smaller
volume $V_c$, the ratio $S_I(E,V_c)/S_I(E,V)$ can be made arbitrarily small by choosing an arbitrarily large value of $V$.

On the other hand, for the equivalent quantum system described in
Sec.~\ref{sec:uncorr_mod}, or more generally, in Appendix
\ref{sec:corr_sys}, the change in entropy may be much smaller given that
the system only ever overlaps of order 50\% with any chosen smaller
volume.  Further analysis requires a definition of entropy, preferably
one that is (a) fully defined in non-equilibrium quantum systems,
(b) generally rises, and (c) corresponds to thermodynamic entropy
for systems in equilibrium.  The ``Observational
entropy"~\cite{safranek2019letter,safranek2019long}, based on a
coarse-graining of Hilbert space, has been shown to satisfy these
properties given a coarse-graining using both position and energy;
we choose this as our test case. (A simpler and more intuitive
argument using another type of Observational entropy coarse-grained
only in position, which however does not directly connect to
thermodynamic entropy, is given in
Appendix~\ref{app:simpleentropyargument}.)

The Observational entropy we employ, denoted $S_{xE}$, entails two sets of coarse-grainings: one that corresponds to measuring coarse-grained position, which for indistinguishable particles is the same as measuring local number of particles, and the second corresponding to measuring total energy.

The positional coarse-graining that will be considered here partitions the Hilbert space into two sectors, one with all the particles confined to the small box, that is $\bx\in X$, and its complement.\footnote{We note that this is a simplified form of $S_{xE}$. In the original form of $S_{xE}$, Ref.~\cite{safranek2019long}, the positional coarse-graining counts the number of particles in each partition, therefore for two partitions, the positional coarse-graining consists of projectors $\C_{\hat{X}}=\{\P_{(N,0)}, \P_{(N-1,1)},\P_{(N-2,2)},\dots\}$, where $\P_{(N_p,0)}$ denotes projector onto subspace corresponding to $N$ particles in the left partition, and $0$ particles in the right partition, and so on. In our simplified version of $S_{xE}$, we have $\P_X:= \P_{(N,0)}$, and all the other projectors are lumped together, into $1-\P_X:= \P_{(N-1,1)}+\P_{(N-2,2)}+\cdots$. We do not expect that this simplification will alter our conclusions, which should be valid even if using the original (non-simplified) $S_{xE}$.} (With the most ``compact" $\ket{\psi}$, the probability of observing the system in $X$ is, as shown in the previous section, of order $1/2$.) We project its position using the projector $\P_X$ defined in \Eq{eq:XProjectorEquivalence}. Then we can write
\[
\label{eq:S_xEequiv-sum_chiEplnp}
S_{xE} := -\sum_{\chi,E} p_{\chi E}\ln\bigg(\frac{p_{\chi E}}{V_{\chi E}}\bigg).
\]
The index $\chi$ can take two values, corresponding to $P_\chi = \P_X$, or $\P_\chi =\P_{X^\perp} = 1- \P_X$, and we have defined
\[
\label{eq:pchiE=braketEPchipsi2}
p_{\chi E} = |\braket{E|\P_\chi}{\psi}|^2,
\]
and
\[\label{eq:Vchi}
V_{\chi E} = \braket{E|\P_\chi}{E}.
\]
$\ket{E}$ has to be an extended state (in the technical sense~\cite{zee2010quantum}, meaning that this energy eigenstate is uniformly distributed over all positions) for a gas, therefore, for any $E$, and independent of $E$, $V_{XE} \sim (V_c/V)^N$, where $V_c$ denotes the physical volume of the region into which we localize, and $V$ denotes physical volume of the full system. For the complement, we assume $V_{X^\perp E} \sim [(V-V_c)/V]^N$.

\Eq{eq:p_X_in_P_X} allows us to write the probability of being in region $X$ as $p_X =\bra{\psi}\hat{P}_X\ket{\psi}$
which is equivalent to
\[
\label{eq:p_chi=psiPXPXpsi}
   p_{\chi} = \bra{\psi}\hat{P}_X \hat{P}_X\ket{\psi} =  \sum_E \braket{\psi}{\hat{P}_X|E}\braket{E|\hat{P}_X}{\psi} = \sum_E p_{\chi E}.
\]

We suppose $\ket{\psi}$ has a probability of $1/2$ that $\bx\in X$; that is, $p_X = 1/2$. Physically this is saying that all of the
particles are inside the volume $V_c := \sum_{\bx \in X} 1 $.

In \Eq{eq:pchiE=braketEPchipsi2} we can expand $\ket{\psi}$ using \Eq{eq:psi_of_zE} and using the fact that the region $X$ has linear dimensions much longer than the thermal length (meaning that the energy spread will be inversely proportional to the box size), we can see that as a function of $E'$, $\bra{E}\hat{P}_X\ket{E'}$ is highly peaked around $E'\approx E$. Therefore we expect that $p_{\chi E} \sim a_E^2$.
As we have assumed, the coefficients are that of ``a pure thermal state," that is $a_E^2 \sim \exp(-\beta E)$. This exponential decay of coefficients means the energies that  significantly contribute come from an energy range centered around the
average energy $\bar{E}$, over some energy shell $\Delta E \ll \bar{E}$. The density of states
$\rho(E)$ is related to the thermodynamic entropy $S_I$,  through $\rho(E) = \exp[S_I(E)]$. Thus
there are
\[
\label{eq:DeltaN=DeltaErho}
   \Delta N = \Delta E \rho(\bar{E})
\]
states that are contributing. Assuming
a flat distribution of $p_{\chi E}$ over this energy window, \Eq{eq:p_chi=psiPXPXpsi} becomes $p_{\chi}=\Delta N p_{\chi E}$. Expressing $p_{\chi E}=p_{\chi}/\Delta N$ and inserting it into \Eq{eq:S_xEequiv-sum_chiEplnp}, we obtain
\[
\begin{split}
   S_{xE} &= -\Delta N \sum_{\chi} \frac{p_{\chi}}{\Delta N} \ln\bigg(\frac{p_{\chi}/\Delta N}{V_{\chi E}}\bigg)\\
   &= -p_X \ln \bigg(\frac{p_X/\Delta N}{(V_c/V)^N}\bigg) - (1-p_X) \ln\bigg(\frac{(1-p_X)/\Delta N}{((V-V_c)/V)^N}\bigg)\\
   &=-p_X \ln p_X-(1-p_X)\ln(1-p_X)+\ln{\Delta N}\\
   &-p_X N \ln (V/V_c)-(1-p_X)N\ln (V/(V-V_c)),
\end{split}
\]
where we have used $V_{XE} \sim (V_c/V)^N$ and $V_{X^\perp E} \sim ((V-V_c)/V)^N$, as explained above (below \Eq{eq:Vchi}).

Writing $\Delta N$ in terms of the thermodynamic entropy and ignoring $\ln \Delta E$ corrections, assuming that the complement is large, $V/(V-V_c)\approx 1$, and ignoring order-1 corrections, we have
\[
   S_{xE} = S_I(E,V)-p_X N\ln(V/V_c).
\]
We know from Appendix \ref{sec:corr_sys} that the size of the $V_c \gg \lambda_T^3$, so that
utilizing~\Eq{eq:ClassicalGasEntropy2} we obtain our final result for entropy,
\[
\label{eq:SxEgtHalfSth}
   S_{xE} > S_I(E,V)-p_X N\ln(V/\lambda_T^3) = (1-p_X) S_I(E,V).
\]

Before we interpret this result, let us take a closer look at the validity of assumptions that we took.

The assumption of a flat distribution of $p_{\chi E}$ over our energy window turns out not to
be a drastic approximation. With a rather general form of this distribution, and including
fluctuations, the main result above is not
altered~\cite{safranek2019long}.

The main questionable assumption in this argument is that \Eq{eq:DeltaN=DeltaErho} is the same for
both values of $\chi$, which means that the wave functions associated with both values of $\chi$ have the same mean energy. We should consider the possibility that $\Delta N$ takes different values in two regions of
Hilbert space because the average energy in these two regions is different.
The probability that all particles in the universe are in the collapsed state is $p_X$, with a probability of $1-p_X$ of being in a highly generic configurations not confined to this collapsed region, i.e. spread throughout the rest of the universe. This then represents a macroscopic
superposition of two very different states.
But if the two elements of this superposition had different temperatures, then the $a_E$'s would no longer look thermal: they would include
two separate separate Gibbs distributions corresponding to two different energy scales. This would be inconsistent with our choice of a single peak for the energy of
the wave function (i.e. $\exp[-\beta E]\exp[S(E)]$).

Thus, returning to the interpretation of result Eq. \eqref{eq:SxEgtHalfSth}, we conclude that when a wave function evolves from a generic pure thermal state, the entropy
can decrease to no less than $1-p_X$ (e.g.~half or $1-\pi^2/16$ in case of real or complex initial wave-function) of the typical thermal value.

On the other hand, if the initial wave function started off being completely confined to the subregion $X$, that is,
$p_X = 1$, then after expanding to fill up the complete volume, it would eventually come arbitrarily close to its initial wave function. In that case, the Observational entropy would behave much as it does in the classical case, so that
$S_{xE}(E,V_c)/S_{xE}(E,V)$ can be made arbitrarily small by choosing an arbitrarily large value of $V$.

\vskip 0.1in

One would also expect that this dichotomy survives even with other definitions of entropy to the degree that entropy is extensive, with contributions  weighted by how much probability is given by the wave function to which physical volume or region of Hilbert space. In this case the compact state would represent rather little entropy, with the probability-1/2 ($1-\pi^2/16$) remainder of the volume representing of order half the original entropy. For example, in a system coarse-grained into volume cells with entanglement entropy between neighboring cells used to quantify entropy, we could expect a comparable result to hold.

Further work~\cite{faiez2019extreme} attempts to verify these results numerically.

\section{Discussion}

Macroscopic effectively-closed systems on terrestrial timescales essentially never significantly decrease their entropy, or evolve away from an equilibrium state.  But on cosmological scales the universe may be, or contain, a closed system that can access indefinitely long timescales in which such evolution would necessarily eventually occur.  This recognition goes back to the time of Boltzmann and has been discussed in a number of papers in recent decades~\cite{aguirre2007eternal,guth2007eternal,banks2003recurrent,dutta2005islands,boddy2015boltzmann,dyson2002disturbing,albrecht2004can}.

This paper demonstrates that there is an intriguing and important
difference in such processes in quantum versus classical physics
regarding whether a many-body system ever evolves so as to ``fit"
into an $M$-dimensional sub-space of its $N$-dimensional accessible
state-space. While a classical non-integrable system fully explores
its accessible phase space so that this will necessarily occur
eventually, in quantum theory the probability of finding a generic
state of the system in the subspace is capped at of order $50\%$,
when $M^2 \ll N$, over all time.

As an example, we consider a subspace where all the particles' positions are in a small box. Our result shows that when starting in a real generic state, the chances of finding all particles in this small box will be less than $50\%$ at any point in time. This is a tight bound in the sense that the chances of finding all of the particles in the box will come arbitrarily close to $50\%$.

It is interesting to note that at first sight, there appears to be a problem with taking the classical limit of our quantum mechanical calculation. Why, in the limit $\hbar \rightarrow 0$, do we not find that this maximum probability approaches $1$? To understand this apparent contradiction, we need to more carefully distinguish the way these two cases are set up.
In the classical ergodic case, we start off with a generically-chosen but precisely determined {\em point} in phase/configuration space and note that it can reach any point consistent with conservation laws; this implies that it can compactify to any volume consistent with these conservation constraints. In the quantum case we start with a generic wave function which gives the probability amplitude of finding particle configurations at any point in Hilbert space. In order to get the classical correspondence from this, we need take a limit where the particles' initial positions become successively more localized as $\hbar \rightarrow 0$, so that in the limit, the particles have precisely-defined and definite positions and momenta. Such a state will be quite unlike a generic state of the original system with $\hbar\neq 0$. Alternatively, we can imagine a classical probability distribution ($\rho$) over phase space that has support over a ``quantum'' of phase space and corresponds to a generic quantum state of the quantized system (see~\cite{safranek2019classical}). In this case there is no reason to think that $\rho$ will condense into a small-volume macrostate.\footnote{Indeed if $rho$ corresponds to an energy eigenstate, then it will be time-independent (and never ``condense''.) So will a sum of such distributions.  Yet any quantum state can be written as a sum of such terms, but {\em can} condense as discussed in this paper; the difference is due to quantum interference.} Thus, when set up in a closely-corresponding way, the difference between the quantum and classical case is not so stark.

Although we did not consider time scales in this problem, it is worth mentioning that these timescales are extremely long. The number of phases of $z_E$ in this problem, $N_z$ is proportional to the exponential of the entropy of the system so that $N_z \propto \exp[S(E)]$. For these all to align at the point where the probability $p_{X}$ achieves its maximum $P_{\max}$
requires an extremely unlikely situation, with a probability proportional to $\exp(-\mathrm{const}\times N_z) \propto \exp[-\exp[S(E)]]$. This means the time for such a collapse is proportional to the inverse of this probability,  $\exp[\exp[S(E)]]$. Taking into account fluctuations discussed in Appendix~\ref{sec:rare_fluctuations}
this will shorten the time but not be expected to change the main exponential factor. Because the entropy depends linearly on the number of particles $N_p$, we expect a time scale that roughly scales as $\exp[\exp[\mathrm{const}\times N_p]]$. The bounds on maximum probabilities that we found also have implications for other quantities. Expressed in a suitable (coarse-grained) entropy measure, the results here indicate that entropy never fluctuates downward by more than $50\%$ in the same limit.

While we have not proven either result in complete generality, our results strongly suggest it is a generic feature of typical quantum many-body systems.  This has several interesting implications.

First, any simple exponential relation between entropy fluctuation magnitude and probability, as suggested by classical fluctuation theorems~\cite{sevick2008fluctuation,crooks1999entropy}, must break down when the entropy fluctuation becomes comparable to the overall entropy. This is studied further in ref.~\cite{faiez2019extreme}.

A second implication is for a quite subtle question: How much {\em information content} is there in a system that has fluctuated from equilibrium? This is related to the paradox represented by Borges' fabled library of all possible books~\cite{borges1962garden}: Does the library contain a vast amount of information (because each book does), or {\em no} information (because as an ensemble the library lends an equal probability to each book)?
 One might square these by arguing that any individual book -- a copy of Hamlet, say -- contains information, but only because it was {\em selected} by some agent; the effort of doing this selection effectively {\em generates} the information associated with that book.  Analogously, any equilibrium system attains many, many distinguishable macrostates and by waiting long enough an observer patiently and repeatedly measuring the system (with unitary evolution between measurements) might eventually find it in essentially any desired macrostate, (generally with exponentially small probability for any given measurement.) One could argue that in this case the information associated with that macrostate is {\em put} into the system by the observer's repeated measurement, and selection of that particular state.

 But the results of this paper add an interesting twist. They
 indicate that for a given coarse-graining into macrostates (Hilbert subspaces), not all equilibrium states are the same. An initially low-entropy state will eventually re-attain low entropy, whilst an initially generic state never will, and must differ in the details of what entropy can be obtained with what probability.\footnote{A followup paper investigates these statistics using numerical methods~\cite{faiez2019extreme}.} As discussed above, this hidden memory appear to operate differently in classical versus quantum systems.

A third implication of the result is for cosmology, where it is widely believed that a low-entropy ``initial'' state of the universe is required to explain the second law and the ``arrows of time''~\cite{carroll2010eternity}.  One possible explanation for this low-entropy state is a large fluctuation away from an overall equilibrium state (e.g.~\cite{albrecht2004can}.)  This explanation encounters various severe objections~\cite{carroll2017boltzmann,dyson2002disturbing,aguirre2012out}.  These boil down to the objection that if one defines macrostates in terms of a limited set of observables (i.e., observables from sufficient to determine the full state of the universe), then such macrostates are much higher entropy than if it were also assumed that entropy was lower in the past, and also lead to very different probabilities for future observations.

Our result that entropy can fall by at most a factor of two arguably adds an additional obstacle to the hypothesis: the analysis here would suggest that this would not be possible. Admittedly, however,
the interpretation of a universal many body wave function is not at all clear.
When constructing a quantum mechanical description of experiments,
there is an observer that is separate from the system of interest,
and that system has associated with it a wave function. Because by definition, there can
be no observer outside of the universe, a universal wave function can no longer be ascribed the same meaning -- and thus nor can ``the entropy of the universe.''

\acknowledgments{
We would like to thank Dana Faiez for useful discussions.
This research was supported by the Foundational Questions Institute (FQXi.org), of which AA is Associate Director, and by the Faggin Presidential Chair Fund.}

\appendix

\section{Correlated systems}
\label{sec:corr_sys}

We will now consider the simplest case where the eigenvectors are not random, that of a
non-degenerate weakly interacting gas in $d$ dimensions of $N_p$ particles in an $L\times L\times L$ box.
We would like to consider this gas at a temperature $T$, and corresponding inverse temperature
$\beta = 1/(k_B T)$.

If we start with a random pure thermal state, so that the coefficients $a_E = |\braket{\psi}{E}|$ are
Gaussian independent random variables with means $\langle a_E^2\rangle = |\braket{\psi}{E}| = \exp(-\beta E)/Z$
Here $Z$ is the partition function. Therefore we can write
\[
a_E = \frac{e^{-\beta E/2}}{\sqrt{Z}} \eta_E,
\]
where $\eta_E$ is a positive random variable and $\langle \eta_E^2\rangle = 1$, so that
$\langle \eta_E\rangle = \sqrt{2/\pi}$. Because we are assuming large $N_p$, the spacing between the
states is very small and we can average the $\eta_E$ over a small energy window that will
still contain many energy eigenvalues, and replace $\eta_E$ by its average value $\sqrt{2/\pi}$.
Therefore \Eq{eq:pX=sum_aE_Ey} becomes
\[
\label{eq:pX=Zstuff}
\begin{split}
&\sqrt{P_{max}} = \sqrt{\frac{2}{\pi}} \frac{Z(\beta/2)}{\sqrt{Z(\beta})} \sum_E \frac{e^{-\beta/2}}{Z(\beta/2)} |\braket{\by_1}{E}|\\
&= \sqrt{\frac{2}{\pi}} \frac{Z(\beta/2)}{\sqrt{Z(\beta})} \langle|\braket{\by_1}{E}|\rangle_{\beta/2},
\end{split}
\]
where the average in the last equality is the canonical average taken at an inverse temperature
$\beta' = \beta/2$.

For any energy $E$ scale, there is a momentum scale, $p$, or wavevector $k=p/\hbar$, that corresponds to
that energy. At inverse temperature $\beta$, there is spatial scale, the thermal wavelength
$\lambda_T$, or thermal wavevector $k_T$, corresponding to the energy scale $k_B T=k_B/\beta$,
\[
\label{eq:lambda_TDef}
\lambda_T := 2\pi/k_T = 2\pi\hbar/p = 2h/\sqrt{2m k_B T}.
\]
The wave function is predominantly made up of wavevectors of order $k_T$ or smaller.

\subsection{Small regions}
\label{sec:SmallRegions}

Let us take the domain $X$ to be a cubical region of width $l$. If $l\ll \lambda_T$,
then the wave function at points inside that region must be almost constant. This fact
will allow us to evaluate $p_X$ for different choices of $\ket{\by_1}$ in order to maximize
$p_X$.

We wish to determine the $\ket{\by_1}$ that will maximize $p_X$. Because $\ket{\by_1}$ can be any superposition
of $\ket{x}$'s for $x\in X$ we try choosing $\ket{\by_1}$ to be
constant for some region inside of $X$. We choose a cube of width $w$,
$X_w \subseteq X$ of width $w$, so that for any point $x\in X_w$,
$\braket{\by_1}{x}$ is constant, but zero outside of this cube. To correctly normalize $\ket{\by_1}$ we take
\[
\braket{\by_1}{x} = \frac{1}{w^{\frac{d N_p}{2}}}
\]
for $x\in X_w$.

Also, $\ket{E}$ is extended throughout all configuration contained in the $L\times L\times L$ box, and for a plane wave $|\braket{x}{E}|$ would be almost constant. If this is a standing wave, this
only changes the normalization by a constant factor of order unity which will make no difference to
our final conclusion. Therefore
\[
\int |\braket{x}{E}|^2  dx^{d N_p}  = L^{d N_p} |\braket{x}{E}|^2 = 1.
\]
Now we can evaluate $\braket{\by_1}{E}$ the limit $l\ll\lambda$,
\[
\braket{\by_1}{E} = \int_X \braket{\by_1}{x}\braket{x}{E} dx^{d N_p} \approx \bigg(\frac{w}{L}\bigg)^{\frac{d N_p}{2}} \eta_E,
\]
where the last factor $\eta_E$ accounts for the fact that the values of $\braket{x}{E}$ have a Gaussian distribution,
and so $\eta_E$ is random and Gaussian with $\langle \eta_E^2\rangle = 1$. We see that $\braket{\by_1}{E}$ is
maximized by choosing $w=l$.

In addition, for a non-degenerate ideal gas, $Z(\beta) = (L/\lambda_T)^{d N_p}$. So using \Eq{eq:pX=Zstuff}, in
the limit of the size of the region much less than the thermal length $l \ll \lambda_T$,
\[
\label{eq:result_small_X}
\sqrt{P_{max}} = \sqrt{\frac{2}{\pi}} \bigg(\frac{2 L}{\lambda_T}\bigg)^{\frac{d N_p}{2}}  \bigg(\frac{l}{L}\bigg)^{\frac{d N_p}{2}} \langle |\eta_E|\rangle
= \frac{2}{\pi} \bigg(\frac{2 l}{\lambda_T}\bigg)^{\frac{d N_p}{2}}.
\]
Therefore in this limit, $P_{max}$ is proportional to the volume of $X$, independent
of system size, but dependent on temperature $T$, and the number of particles $N_p$.

%Note that \Eq{eq:result_small_X} becomes of order unity when $l = \lambda_T/2$. At that point, the
%equation is invalid, but it suggests a crossover to

\subsection{Larger regions}
\label{LargeRegions}

For larger regions, $X$, the evaluation of \Eq{eq:pX=Zstuff} becomes more
difficult, because we must find the correct basis vector $\ket{\by_1}$ according to the prescription
of Sec. \ref{sec:maximization}.  However in the opposite limit to what we just considered,
that is for $l$ is sufficiently large, we will now argue
that this system becomes closely related to the case of uncorrelated eigenvectors analyzed in Sec. \ref{sec:uncorr_mod}.
A technical problem is that we had previously considered a finite dimensional Hilbert space, whereas
now this space is infinite dimensional. We can handle this by
$a_E = 0$ above some cutoff energy $E_c$. Because the $a_E$ decrease exponentially, such a cutoff will
have no effect in the limit as $E_x\rightarrow\infty$.

Because the very large energy eigenvectors contribute negligibly, it is inconvenient to use
use the position basis, but instead we choose to use a Wannier basis~\cite{WannierRevModPhys.34.645} to represent coarse grained
position.

The transformation into this Wannier basis can be done in two steps. The first is to lay down
lattice points separated by some distance $D$, say on a cubic lattice. We will take $D \gg
\lambda_T$.
For example in two dimensions, we can take ${\bf R} =j_1 D\hat{x}+j_2 D\hat{y}$, where $j_1$ and $j_2$ are integers.
Then we consider
single particle momentum eigenstates $\ket{{\bf K}}\propto\int\exp(i{\bf K}\cdot{\bf r})\ket{{\bf r}}d^dr)$
and write this as a Bloch wave function by reindexing $\bf k$ in terms of crystal momentum and
band index $\braket{{\bf K}}{{\bf r}} = \braket{{\bf k},n}{\bf r}$, where $\bf k$ can be confined to the
first Brillouin zone~\cite{AshcroftAndMermin} and $n$ is the band index.

Thus the Wannier basis contains two indices, the position of lattice points, $\bf R$,
and an additional integer index, $n$, representing the band, associated with each lattice points.  Utilizing an arbitrary
(real) phase  function $\theta({\bf k})$ we can write
\[
\ket{{\bf R},n} = \bigg(\frac{L}{2\pi}\bigg)^d\int e^{i\theta({\bf k})} \ket{{\bf k},n} \exp(-i{\bf k}\cdot{\bf R}) d^d k,
\]
where the integral is taken over the first Brillouin zone.
This basis is orthonormal and complete,
and the $\braket{{\bf R},n}{r}$ can be shown to be of the form $\phi_n({\bf r}-{\bf R})$
where $\phi_n$ is localized for appropriate choice of $\theta({\bf k})$. Even with the choice $\theta=0$,
the probabilities associated with those states decay for large distance $r$, have a power law envelope
proportional to $1/x^2$ along every axis $x$, leading to confinement of probability to a local
region around a lattice point.

To express $p_X$ in this basis, we can write for a single particle
\[
\ket{\psi} = \sum_{\bR,n} \braket{\bR,n}{\psi}\ket{\bR,n},
\]
and in this basis,
\[
\begin{split}
&p_X = \int_X |\braket{r}{\psi}|^2 d^d r \\
&= \sum_{\bR,n} \sum_{\bR',n'} \braket{\bR,n}{\psi} \braket{\psi'}{\bR',n} \int_X \braket{\bR',n'}{\bR,n} d^d r\\
&\approx \sum_{\bR\in X,n} |\braket{\bR,n}{\psi}|^2.
\end{split}
\]
The last line uses the orthonormality of $\ket{\bR,n}$, if the integration is over all $\bR$. Because
the integration here is confined to the region $X$, the last line is an approximation. Since
the Wannier functions can be chosen to be well localized, it should be a good one for box widths
much greater than the lattice spacing,  $l\gg D$.

For $N_p$ particles,
the corresponding generalization of such states is
$\ket{\{{\bf R}_i,n_i\}_i} := \ket{{\bf R_1},n_1}\otimes\ket{{\bf R_2},n_2}\otimes\dots\otimes\ket{{\bf R}_{N_p},n_{N_p}}$, and $X$ denotes a region in $d N_p$ dimensional space, $X = (X_1, X_2,\dots,X_{N_p})$, where $X_i$ is a $d$
dimensional cubical region of width $l$.
Therefore, we can equivalently ask for the probability
\[
\label{eq:prod_sum_sum_R_n_psi_sq}
p_X = \prod_{i=1}^{N_p}\bigg[\sum_{n_i}\sum_{R_i\in X_i}|\braket{\{{\bf R}_j,n_j\}_j}{\psi}|^2\bigg].
\]
We can write \Eq{eq:prod_sum_sum_R_n_psi_sq} in terms of energy eigenstates
\[
p_X = \prod_{i=1}^{N_p}\bigg[\sum_{n_i}\sum_{R_i\in X_i}|\sum_E \braket{E}{\psi} \braket{\{{\bf R}_j,n_j\}_j}{E}|^2\bigg].
\]
An eigenstate of a weakly interacting gas will be well approximated by a sum of plane wave, each
plane wave of the form $\exp(i\sum_i {\bf k}_i\cdot{\bf r}_i)$.
However due to scattering, the wave function will become uncorrelated beyond
the scattering length $\xi$. We will assume that $\xi \gg D$, the lattice spacing of
the Wannier states. We already assumed that $D\gg\lambda_T$ and so this value of $\xi$ implies
weak scattering. Because a Wannier state for one particle only has contributions from
a single band index $n$, and the scattering is taken to be weak,
an energy eigenstate is still well approximated to have contributions only from a single band index $n$.
We can also separate out the product and summations to write
\[
\prod_{k=1}^{N_p}\bigg[\sum_{n_k}\sum_{R_i\in X_i}\bigg]
=\bigg[\prod_{k=1}^{N_p}\sum_{n_k}\bigg]\bigg[\prod_{i=1}^{N_p}\sum_{R_i\in X_i}\bigg].
\]
This means that we can write
\[
p_X = \bigg[\prod_{k=1}^{N_p}\sum_{n_k}\bigg]\bigg[\prod_{i=1}^{N_p}\sum_{R_i\in X_i}\bigg]\bigg\rvert\sum_{E\in E(\{n_i\}_i)} \braket{E}{\psi} \braket{\{{\bf R}_j,n_j\}_j}{E}\bigg\rvert^2.
\]
The inner sum over energy is confined to the specific bands that are indexed in the outer summation. As
in Sec. \ref{sec:uncorr_mod}, we denote $\braket{E}{\psi} := a_E z_E$.
Therefore when taking the maximum of $p_X$ over all values of $z_E$, we can maximize each
combination of bands $\{n_i\}_i)$ separately,
\[
\label{eq:P_max=prod_max_prod_sum_R_n}
\begin{split}
P_{max}& = \prod_{k=1}^{N_p}\sum_{n_k}\\
&\max_{z_E, E\in E(\{n_i\})} \prod_{i=1}^{N_p}\sum_{R_i\in X_i}\bigg\rvert\sum_{E\in E(\{n_i\})} a_E z_E \braket{\{{\bf R}_j,n_j\}_j}{E}\bigg\rvert^2.
\end{split}
\]

Now consider the special case where $a_E=0$ unless $E\in E(\{n'_i\}_i)$, where the $\{n'_i\}_i$ are
some specific choice of band indices. If the energy eigenstates are within these bands, then we
choose $\langle a_E^2\rangle$ to be constant. For a single particle, the number of states within a band is
$(L/2\pi)^d$, and for $N_p$ particles, the number of states is $N_n = (L/2\pi)^{N_p d}$.
Therefore $\langle a_E^2\rangle  = 1/N_n$.
Every particle has states inside only one band, and for that band $n$, the Wannier
states $\ket{\bR,n}$ form a complete orthonormal set. In that case, we have precisely the situation
studied in \ref{sec:uncorr_mod}, where we found that $P_{max} = 1/2$ for $M^2/N \ll 1$. In this case,
this condition is satisfied when
\[
\frac{M^2}{N} = \bigg(\frac{(l/D)^2}{L/D}\bigg)^{N_p d} \ll 1.
\]
The value of $D$ here was chosen to be arbitrary with $D\gg\lambda_T$. This means that we expect
that a more stringent criterion for the subspace size $l$ is
\[
\bigg(\frac{l^2}{L\lambda_T}\bigg)^{N_p d} \ll 1.
\]
Because $N_p$ is taken to be very large, this will be satisfied
for $l < \sqrt{\lambda_T L}-\epsilon$, where $\epsilon\rightarrow 0$ as $N_p\rightarrow\infty$.
For the argument to apply, the eigenvectors should have random statistics in the Wannier basis.
There can still be short range correlations, but the system size should be larger than this
correlation length. We therefore should add the condition that the box size is much
greater than the scattering length, $l \gg \xi$.

Now consider the thermal case for the coefficients $a_E$. For the case that we are considering,
$d \gg \lambda_T$, $\langle a_E^2\rangle$ is almost constant within one band.
Therefore by rescaling the $a_E$ appropriately, \Eq{eq:P_max=prod_max_prod_sum_R_n} becomes
\[
\begin{split}
\label{eq:P_max=prod_sum_asq_ov_N_n_1_2}
P_{max}& = \prod_{k=1}^{N_p}\sum_{n_k} \frac{\langle a^2_{E(\{n_i\}_i)}\rangle}{1/N_n}\frac{1}{2}.
\end{split}
\]
Here the notation $E(\{n_i\}_i)$ means the minimum energy of a particular set of bands. As mentioned
above  $\langle a^2_E\rangle$ is taken to be constant for all $k$ values of these bands. This allows
us to rescale $\langle a^2_E\rangle$ and identify the maximization problem with the special case
analyzed above.
By breaking up the different energy levels into their bands, and then particular energy state in a band,
we have that
\[
1 =\sum_E \langle a^2_E\rangle = \bigg[\prod_{k=1}^{N_p}\sum_{n_k}\bigg] \sum_{E\in E(\{n_i\})}\langle a^2_E\rangle.
\]
Now with the same assumption of small variation of $\langle a^2_E\rangle$ inside a single band,
\[
1 =\prod_{k=1}^{N_p}\sum_{n_k} \langle a^2_{E(\{n_i\}_i)}\rangle\sum_{E\in E(\{n_i\})} 1
=\prod_{k=1}^{N_p}\sum_{n_k} \langle a^2_{E(\{n_i\}_i)}\rangle N_n.
\]
Therefore \Eq{eq:P_max=prod_sum_asq_ov_N_n_1_2} becomes
\[
 P_{max} = 1/2
\]
for the condition given above, essentially that
$l \ll \sqrt{\lambda_T L}$ as $N_p \rightarrow \infty$. We also required $l \gg \xi$
for our argument to work. Above the threshold $l < \sqrt{\lambda_T L}$, $M^2$ rises very
sharply and according to the scaling that we had previously found, $P_{max}(M^2/N)$, we expect the
probability to rapidly rise to a number close to $1$.

\section{Rare Fluctuations}
\label{sec:rare_fluctuations}

We can extend the analysis of the maximum probability $p_X$ for the uncorrelated eigenvector model
of Sec.~\ref{sec:uncorr_mod}, to ask what is the distribution of rare
fluctuations in a region in the quantity $p_X$, as defined in~\Eq{eq:p_X=sum_p_x}. That is, we would like to calculate
\[
\Pc(p) := \langle \delta(p_X(t) - p)\rangle_t,
\]
where the angular brackets denote an infinite time average. $\Pc(p) dp$ is the probability of encountering
the system with $p_X$ between $p$ and $p+dp$.
The time dependence in $p_X$ comes in through
the coefficients in \Eq{eq:def_psi_t}, where $c_E(t)$ has a time dependence
$c_E(t) = \exp(i\theta(t)) c_E(0)$, and the energy phase angle $\theta(t) = E t$. Therefore the
for long times, all phase angles will be uniformly covered and therefore we can equivalently
average over phase angles
\[
\Pc(p) = \langle \delta(p_X(\{\theta_E\}_E)- p)\rangle_{\{\theta_E\}_E}.
\]

We can make an analogy with statistical mechanics, and
think of $p_X$ as a fake ``Hamiltonian" that depends on the phase angles, $H_X:= -p_X$, and $\Pc(p)$
is the probability density of phase angles. Therefore $\Pc(p)$ is related to the entropy as a function of energy
because
\[
\begin{split}
\Pc(p) &= \frac{\int \delta(p_X(\{\theta_E\}_E)- p) \prod_E d\theta_E}{\int  \prod_E d\theta_E}\\
&=\Big(\frac{1}{2\pi}\Big)^N\int \delta(p_X(\{\theta_E\}_E)- p) \prod_E d\theta_E\\ &=\Big(\frac{1}{2\pi}\Big)^N \Omega(p)
=\Big(\frac{1}{2\pi}\Big)^N e^{S(p)}.
\end{split}
\]
Here $\Omega$ is the phase space volume of the region on the surface $p_X = p$, which is related to the entropy
$S(p)$ and in this analogy~\cite{ma1985statistical}, we have sensibly set Boltzmann's constant to unity.

The ``Hamiltonian", is actually the same as that for a classical $xy$ spin system. We write the
probability $p_\bx$ using the unit magnitude complex numbers $z_E$ introduced in  \Eq{eq:psi_of_zE}
as
\[
\label{eq:APP:p_x_z_E=sum_a_E_z_E}
\begin{split}
   p_\bx(\{z_E\}_E) &= \big|\sum_E a_E z_E \braket{\bx}{E}\big|^2\\
   &= \sum_{E,E'} a_E a_{E'} z_E z^*_{E'} \sum_{\bx\in X} \braket{\bx}{E}\braket{E'}{\bx}\\
   &= \sum_{E,E'} z_E J_{E E'} z^*_{E'} .
\end{split}
\]
Now we
write the $z_E$  as two dimensional unit vectors
vectors $\vec{s}_E$, where the real and imaginary parts of $z_E$ correspond respectively to the $x$ and $y$
components of $\vec{s}_E$. Then, combining
\Eq{eq:p_x_z_E=sum_a_E_z_E} and \Eq{eq:p_X=sum_p_x}, we can write

\[
\label{eq:H_X=sumJsdots}
H_X = -\sum_{E,E'} J_{EE'}\vec{s}_E\cdot\vec{s}_{E'},
\]
where the coupling was already used in in \eqref{eq:JEE=sumaaExxE}, namely
\[
\label{eq:APP:JEE=sumaaExxE}
J_{EE'} = \sum_{x\in X} a_E a_{E'}\braket{E}{x}\braket{x}{E'}.
\]
This is closely related to a neural network model for associative memory, the
Hopfield model~\cite{hopfield1982neural}, but there the Ising spins are used rather than $xy$ spins.
A slightly different version of the Hopfield model with $xy$ spins has been recently studied~\cite{kimoto2013continuous}.

We can get the low energy behavior of this model, by expanding it in the usual way for low lying
excitations, up to quadratic order in the deviations, $\delta\theta_E$, in
the phase angles from their ground state values,
\[
H_X(\{\theta_E\}_E) \approx H^{min}_X + \frac{1}{2}\sum_{EE'} \delta\theta_E M_{EE'} \delta\theta_{E'}.
\]
In general from numerical work, the minima are nearly degenerate, and therefore the volume dependence as
a function of energy is given by the volume of a hypersphere of radius proportional to $\sqrt{H_X-H^{min}_X}$. This argument breaks down when the quadratic approximation breaks down which will certainly be the case for high enough energies, but if the ``energies" are close enough to the ground state, this should give a reasonable approximation.
Therefore
\[
\Omega(H_X) \propto (H_X-H^{min}_X)^{\frac{N-2}{2}}.
\]
Translating this back into our original variables $\Pc{p_X}$, we have~\cite{ma1985statistical}
\[
\Pc(p) \propto \Omega(p) \propto (P_{max}-p)^{\frac{N-2}{2}}.
\]
This demonstrates that the probability of finding lower values than $P_{max}$ rises extremely
rapidly, as a power law depending on the dimension of the Hilbert space.

\section{Simplified argument for entropy downward fluctuation}\label{app:simpleentropyargument}

Consider Observational entropy that employs just the positional coarse-graining,
\[
S_{x} := -\sum_{\chi} p_{\chi}\ln\bigg(\frac{p_{\chi}}{V_{\chi}}\bigg),
\]
where index $\chi$ can take two values, corresponding to $P_\chi = \P_X$, or $\P_\chi = 1- \P_X:= \P_{\neg X}$. We can write
\[
S_{x} = - p_{X}\ln {p_{X}}- p_{\neg X}\ln {p_{\neg X}}+p_{X}{V_{x}}+p_{\neg X}{V_{\neg X}}.
\]

With the most ``compact" $\ket{\psi}$,
the probability of observing the system in $X$ is $p_{X}=1/2$. Then we have
\[
\begin{split}
S_{x}^{\mathrm{(compact)}} &= - \tfrac{1}{2}\ln \tfrac{1}{2}- \tfrac{1}{2}\ln \tfrac{1}{2}+\tfrac{1}{2}\ln {V_{X}}+\tfrac{1}{2}\ln {V_{\neg X}}\\
&\approx \tfrac{1}{2}\ln {V_{\neg X}}\approx \tfrac{1}{2}\ln \dim \HS,
\end{split}
\]
since subspace $X$ is much smaller than the rest of the Hilbert space (describing the rest of the ``universe''), $M=V_X=\dim X\ll\dim \neg X= V_{\neg X}$. Additionally, the rest of the Hilbert space is almost the same size as the entire Hilbert space, $V_{\neg X}\approx \dim \HS=N$.

The wave function of an initial random state is almost entirely contained in the complement $\neg X$, $p_{\neg X}=1$, therefore
\[
S_{x}^{\mathrm{(initial)}}= \ln V_{\neg X}\approx \ln \dim \HS.
\]

Together, we have
\[
\frac{S_{x}^{\mathrm{(compact)}}}{S_{x}^{\mathrm{(initial)}}}=\frac{1}{2}.
\]

\section{Unitary and global maximization equivalence}
\label{sec:UnitaryGlobalMaxEquiv}

We first write the results in Sec. \ref{sec:maximization} directly in terms of the phases where
the probability will be extremized. In other words we want to write the right hand side of, Eq.~\eqref{eq:maxzE}
namely
\[
\label{eq:APP:maxzE}
\tilde{z}_E := \frac{\braket{E}{\by_1(U)}}{|\braket{E}{\by_1(U)}|}.
\]
in terms of $z_E$ as well.

To do this, we use Eq. \eqref{eq:y1=proj_psi}, namely
\[
 \label{eq:APP:y1=proj_psi}
 \ket{\by_1}:=\frac{\hat{P}_X\ket{\psi}}{||\hat{P}_X\ket{\psi}||}.
\]
to see that
\[
   \label{eq:zEproptosumExxpsi}
\begin{split}
   \arg(\tilde{z}_E) &= \arg\big(\sum_{\bx} \braket{E}{\bx} \braket{\bx}{\by_1}\big)
   = \arg\Big(
   \sum_{\bx} \braket{E}{\bx} \frac{\braket{\vec{x}|\hat{P}_X}{\psi}}{||\hat{P}_X\ket{\psi}||}\Big)\\
   &= \arg\big(\sum_{\bx\in X} \braket{E}{\bx} \braket{\bx}{\psi}\big),
\end{split}
\]
where $\arg$ denotes the complex phase, $z=|z|e^{i\arg z}$.

Now we take an inner product in Eq. \eqref{eq:psi_of_zE},
\[
   \braket{\bx}{\psi(\{z_E\}_E)} = \sum_E a_E z_E \braket{\bx}{E}.
\]
so that the last equality in Eq. \eqref{eq:zEproptosumExxpsi} becomes

\[
   \arg(\tilde{z}_E) = \arg\big(\sum_{E'} a_E a_{E'} \tilde{z}_{E'} \sum_{\bx\in X} \braket{E}{\bx}\braket{\bx'}{E'}\big)
\]

Defining $J_{E E'} $ as in Eq. \eqref{eq:JEE=sumaaExxE}
allows us to more simply write
\[
   \label{eq:APP:z_EproptosumJz}
   \arg(\tilde{z}_E) = \arg\big(\sum_{E'} J_{E E'} \tilde{z}_{E'}\big)
\]

Now we compare this directly to the condition that we have to extremize the probability. This probability can be written as in Eq.
\eqref{eq:APP:p_x_z_E=sum_a_E_z_E}.
To find the extrema, we differentiate its right hand side with respect to all of the $z^*_E$ subject to the constraint
that $z_E z^*_E = 1$. We can include these constraints using the method Lagrange multipliers by adding a term
\[
-\sum_E \lambda_E z_E z^*_E
\]
in the right hand side of Eq. \eqref{eq:APP:p_x_z_E=sum_a_E_z_E}, where the $\{\lambda_E\}_E \in \mathbb{R}$ are the
Lagrange multipliers. Differentiating and setting this
equal to zero, we obtain Eq.~\eqref{eq:APP:z_EproptosumJz} above. This could be equivalently be
performed
using the spin variables  $\vec{s}_E$ used in Appendix~\ref{sec:rare_fluctuations}.

This shows that all extrema are captured by the ansatz of Sec.  \ref{sec:maximization}.

\section{Iterative procedure to find maximum}
\label{IterativeProcedureForMax}

Formulating the maximization problem as a minimization problem of a spin Hamiltonian Eq. \eqref{eq:H_X=sumJsdots},
we are looking for spin configurations that minimize this Hamiltonian. To derive a method for iteratively solving this equation
for maxima, we can introduce
an iterative procedure based on relaxational dynamics of this spin system. We introduce fictitious
time variable $t$ and consider the evolution of the spin variables over time in a way that will lead
to energy minimization. This is a special case of the Landau Lifshitz Gilbert equation
~\cite{eriksson2017atomistic},
\[
   \label{eq:SpinDynamics}
   \frac{d \vec{s}_E}{dt} = -\gamma \vec{s}_E\times(\vec{s}_E\times\frac{\partial H}{\partial \vec{s}_E})
\]
where $\gamma$ is some positive damping factor (that for our purposes can be time dependent).
One can interpret the motion of these spins as being due to forces
\[
   \label{eq:fE=dHds}
   \vec{f}_E := -\frac{\partial H}{\partial \vec{s}_E} = \sum_E J_{E E'} \vec{s}_{E'}
\]
It follows that
\[
 \vec{s}_E\cdot\frac{d \vec{s}_E}{dt} = 0
\]
by using the right hand side of Eq. \eqref{eq:SpinDynamics} and the perpendicular nature of cross products.

We can also compute the rate of energy change using standard cross product identities
\[
\begin{split}
   \frac{d H}{d t} &= \sum_E \frac{\partial H}{\partial \vec{s}_E}\cdot \frac{d \vec{s}_E}{dt}\\
   &= - \gamma \sum_E \vec{f}_E\cdot(\vec{f}_E-(\vec{f}_E\cdot\vec{s}_E)\vec{s}_E)\\
   &= - \gamma \sum_E f_E^2 - (\vec{f}_E\cdot\vec{s}_E)^2 \le 0 ,
\end{split}
\]
showing that these dynamics continually lower the energy until the spins cease moving and are therefore aligned with the
forces. The condition that these two sets of vectors are aligned, is equivalent to Eq. \eqref{eq:z_EproptosumJz} as can be
seen from translating this to spin notation
\[
   \label{sEproptosumJs}
   \vec{s}_E \propto \vec{f}_E = \sum_{E'} J_{E E'} \vec{s}_{E'}
\]
and using Eq. \eqref{eq:fE=dHds}.

To implement this equation numerically, we can regard the forces at time $t$ to be a function of the spins
$\vec{f}_E(\{\vec{s}_E\}_E$ at the same time, and iterate over time steps $\Delta t$
\[
   \label{eq:sEtDeltat=foverabsf}
   \vec{s}_E(t+\Delta t) = \frac{\vec{f}_E(\{\vec{s(t)}_E\}_E}{|\vec{f}_E(\{\vec{s(t)}_E\}_E|}
\]
with the $\vec{f}_E$ defined in \eqref{eq:fE=dHds}.
The fixed points of this equation are the same as Eq. \eqref{sEproptosumJs}, which as we saw, gives the extrema of the
probability. We can understand what happens when spins $\{\vec{s}_0\}$ are close to a fixed point.

Consider the spins at time $t$, $\{\vec{s}_t\}$ by writing the corresponding forces and dividing them into components parallel
and perpendicular to $\{\vec{s}_t\}$,
\[
   \vec{f} = \vec{f_\parallel}  + \vec{f_\bot} .
\]
Substituting this into Eq. \eqref{eq:sEtDeltat=foverabsf} (and dropping the $E$ subscript for
clarity)
\[
\begin{split}
   \vec{s}(t+\Delta t) &= \frac{\vec{f_\parallel}}{f_\parallel} + \frac{\vec{f_\bot}}{f} + O(f_\bot^2)\\
   &=  \vec{s}(t) + + \frac{\vec{f_\bot}}{f} + O(f_\bot^2)
\end{split}
\]

Noting the $\vec{f_\bot} = \vec{s}\times(\vec{s}\times\vec{f})$ we see that for small $f_\bot$, and
choice of the appropriate $\gamma$, this equation is equivalent to a discretized version of Eq. \eqref{eq:SpinDynamics} up to
second order correction in $f_\bot$. When sufficiently close to a minimum of $H$, $f_\bot$ becomes
arbitrarily small and these second order corrections become negligible. Therefore this iterative procedure will
lead to a maximization of $p_X$.

\bibliography{probability_limits}
\end{document}